\documentclass{aastex}
\usepackage{emulateapj5,apjfonts,psfig}

\def\asca{{\sl ASCA }}

\def\ros{{\sl ROSAT }}
\def\gin{{\sl GINGA }}

\def\ein{{\sl Einstein }}

\def\chandra{{\sl Chandra }}
\def\euve{{\sl EUVE }}
\def\xmm{{\sl XMM-Newton }}
\def\ergsec{\hbox{erg s$^{-1}$ }}
\def\ergcm{\hbox{erg cm$^{-2}$ s$^{-1}$ }}

\def\Msun{$M_{\odot}$ }
\def\Rsun{$R_{\odot}$ }

\def\it{\sl}

\slugcomment{accepted for publication in The Astrophysical Main Journal}

\shorttitle{X-RAY MODELING OF HOT STARS}

\shortauthors{SCHULZ et al.}

\received{2002 December 2}

\accepted{2003 May 30}
\begin{document}

\title{X-ray Modeling of Very Young Early Type Stars in the Orion Trapezium: Signatures
of Magnetically Confined Plasmas and Evolutionary Implications}
\author{
N.S. Schulz,\altaffilmark{1}
C.Canizares,\altaffilmark{1}
D.Huenemoerder,\altaffilmark{1}
K. Tibbets\altaffilmark{1}
}
\altaffiltext{1} {Center for Space Research, Massachusetts Institute of Technology, Cambridge MA 02139, USA}

\begin{abstract}
The Orion Trapezium is one of the youngest and closest star forming regions within our Galaxy.
With a dynamic age of $\sim 3\times 10^5$ yr it harbors a number of very young hot stars, which
likely are on the zero-age main sequence (ZAMS). We analyzed high resolution X-ray spectra in the
wavelength range of 1.5 -- 25 \AA~ of three of its X-ray brightest members ($\Theta^1$ Ori A, C and E) 
obtained with the 
High Energy Transmission Grating Spectrometer (HETGS) on-board the Chandra X-ray Observatory.
We measured X-ray emission lines, calculated differential emission
measure distributions (DEMs), and fitted broad-band models to the spectra. The spectra from all
three stars are very rich in emission lines, specifically from highly ionized Fe which includes emission
from Fe XVII to Fe XXV ions. A complete line list is included.  
This is a mere effect of high temperatures rather than an overabundance of Fe, which
in fact turns out to be underabundant in all three Trapezium members. Similarly there
is a significant underabundance in Ne and O as well, whereas Mg, Si, S, Ar, and Ca appear close to solar. 
The DEM derived from over 80 emission lines in the spectrum of
$\Theta^1$ Ori C indicates three peaks located at 7.9 MK, 25 MK, and 66 MK. The emission measure varies over
the 15.4 day wind period of the star. For the two phases observed, the low temperature emission
remains stable, while the high temperature emission shows significant differences. The line widths
seem to show a similar bifurcation, where we resolve some of the soft X-ray lines with 
velocities up to 850 km s$^{-1}$ (all widths are stated as HWHM), whereas the bulk of the lines 
remain unresolved with a 
confidence limit of 110 km s$^{-1}$. The
broad band spectra of the other two stars can be fitted with several collisionally ionized plasma
model components within a temperature range of
4.3 -- 46.8 MK for $\Theta^1$ Ori E and 4.8 -- 42.7 MK for $\Theta^1$ Ori A. 
The high temperature emissivity contributes over 70$\%$ to the total X-ray flux.
None of the lines are resolved for $\Theta^1$ Ori A and E with a confidence limit of 160 km s$^{-1}$. 
The influence of the strong UV radiation field on the forbidden line in the He-like triplets allows us to set 
an upper limit on distance of the line emitting region from the photosphere.  
The bulk of the X-ray emission cannot be produced by shock instabilities in a radiation 
driven wind and are likely the result of magnetic confinement in all three stars. Although confinement models
cannot explain all the results, the resemblance of the unresolved lines and the of the DEM  
with recent observations of active coronae in II Peg and AR Lac during flares is quite obvious. 
Thus we speculate that the X-ray production mechanism in these stars is similar with the difference
that the Orion stars maybe in a state of almost continuous flaring driven by the wind.
We clearly rule out major effects due to X-rays from a possible companion.
The fact that all three stars appear to be magnetic and are near zero age on the main sequence 
also raises the issue whether the Orion stars are simply different or 
whether young massive stars enter the main sequence carrying significant magnetic fields. 
The ratio log $L_x/L_{bol}$ using the `wind' component of the spectrum is -7
for the Trapezium consistent with the expectation from O-stars. This suggests  
that massive ZAMS stars generate their X-ray luminosities like normal O stars and magnetic confinement provides
an additional source of X-rays. 
\end{abstract}


\section{Introduction}

Since the discovery of X-ray emission from massive early type stars more than two decades ago 
(Seward et al. 1979, Harnden et al. 1979) it is an ongoing quest to explain its
origins and to develop physical models that consistently predict its characteristics. 
Although to date no definitive models for the production of X-rays in stellar winds exist,
it is widely established  that X-rays are produced by shocks forming from instabilities within a
radiatively driven wind (Lucy $\&$ White 1980). This phenomenological model has been revised
and expanded throughout the years. Lucy (1982) showed that these shocks can exist well out
into the terminal flow overcoming the attenuation problem of the previous model. Owocki, Castor,
and Rybicki (1988) extended the model in that they showed that reverse shocks are much stronger
than forward shocks in an high velocity gas at low densities, deduced from P Cygni absorption
in UV resonance line profiles (Puls et al. 1993, Hillier et al. 1993). 
These models were able to successfully explain the mass loss from a hot luminous star in the UV domain
as well as the soft X-ray spectral temperatures, but to date cannot correctly predict observed X-ray fluxes. 
For example,  Feldmeier, Puls $\&$ Pauldrach (1997) see the possibility of mutual collisions of dense gas shells
in the outer wind to produce stronger shocks. 
Claims that the X-ray emission could originate from coronal gas near the stellar photosphere  
were soon considered unlikely due to the lack of 
soft X-ray absorption edges (Cassinelli $\&$ Swank 1983) as well as the absence
of coronal emission lines in optical spectra (Nordsieck, Cassinelli, $\&$ Anderson 1981). 

That all O and early type B stars are strong stellar X-ray emitters is now a well established fact
thanks to relentless observations with \ein and \ros 
(Pallavicini et al. 1981, Chlebowski et al. 1989, Bergh\"ofer et al. 1994,
Cassinelli et al. 1994). Typical X-ray luminosities are of the order of $10^{32}$ \ergsec for
O-stars and  $10^{30.5}$-$10^{31.5}$ \ergsec for B-stars 
(Bergh\"ofer, Schmitt, and Cassinelli 1996),
while many low-mass (late type) PMS stars radiate at orders of magnitude lower luminosity and 
only the peak of their luminosity function reaches $10^{31}$ \ergsec (Feigelson and Montmerle 1999).
Due to the lack of spectral resolving power, however, many results from
\ein and \ros were based on statistical properties of a large sample
of Stars (Chlebowski et al. 1989, Bergh\"ofer, Schmitt, and Cassinelli 1996). 
Among these results were that the X-ray luminosity in early type stars typically scales with
the bolometric luminosity with log (L$_x$/L$_{bol}$) = -7 and with a few exceptions
as high as -5. 

In the first very detailed spectral analysis using the ROSAT PSPC, Hillier et al. (1993)
fitted spectra of $\zeta$ Pup with NLTE models under the assumptions that
the X-rays arise from shocks distributed throughout the wind and that recombination occurs in the outer
regions of the stellar wind. The best fits predicted two temperatures of log T(K) $\sim$ 6.2 to 6.7 with shock velocities
around 500 km s$^{-1}$. Based on this approach, Feldmeier et al. (1997) added the assumption that the X-rays originate
from adiabatically expanding cooling zones behind shock fronts and described the spectra with post shock temperature
and a volume filling factor. These results were also compared to results from Cohen et al. (1996), who used
\ros and \euve data to constrain high-temperature emission models in the analysis of the B-giant, $_p3.epsilon$ CMa.
A continuous temperature distribution was inferred over single or even two
temperature models. The result of that comparison remained inconclusive, since both views appeared
indistinguishable in the spectra.  
Despite the success of the wind shock models, several unanswered issues remain from the 
\ein, \ros, and \asca era, which seem
to be quite in contrast to this model (see also below). One issue concerns the unusually hard X-ray spectra of the
B0.2 V star $\tau$ Sco observed with \asca (Cohen, Cassinelli, $\&$ Waldron 1997),
of $\lambda$ Ori (Corcoran et al. 1994), and of
\ros spectra of stars later than of B2 type (Cohen, Cassinelli, $\&$ MacFarlane 1997).

With the availability of high resolution spectra from \chandra and \xmm the spectral situation 
became much more complex and confusing. The first published high resolution X-ray spectrum 
of the Orion Trapezium star $\Theta^1$ Ori C showed extreme temperatures and symmetric lines
(Schulz et al. 2000 -- paper~I). These properties are not expected from shocked material 
near or beyond regions where the wind reached its terminal velocity. Based on a soft X-ray spectrum 
and symmetric emission lines from $\zeta$ Ori, Waldron and Cassinelli (2001) argued that the emitting
plasma originates likely near the photosphere. 
Highly resolved spectra from $\zeta$ Pup with \xmm (Kahn et al. 2001) and \chandra (Cassinelli et al. 2001)
finally showed some expected emission characteristics, i.e. moderate temperatures of 5 to 10 MK, blue shifted and
asymmetric lines. Such X-ray line profiles (Ignace 2001, Owocki and Cohen 2001) are
significant characteristics of attenuated plasma moving X-ray emitting plasma. Schulz et al. (2001a) and 
Schulz et al. (2002) report on similar evidence from line profiles in HD 206267 and $\iota$ Ori, respectively.

The fact that the Orion Trapezium star $\Theta^1$ Ori C shows such rather strange X-ray characteristics
may not come as too much of a surprise, since this star was already known to be of rather 
peculiar nature (Stahl et al. 1995, Gagne et al. 1997). Babel and Montmerle (1997) proposed an aligned
magnetic rotator model. The Orion Trapezium region has always been quite difficult to observe
prior to \chandra simply for the fact that its constituent members could never be spatially resolved.
The \ros HRI (Gagne et al. 1995) provided better spatial resolution, but no
spectral information. Yamauchi and Koyama (1993) observed hard X-rays from the 
Orion Nebula region with \gin that did not rule out the possibility of 2-3 keV X-rays from 
$\Theta^1$ Ori C, but focused more on possible hard extended emission
within the Nebula. This issue was further studied with \asca (Yamauchi et al. 1996). Highly resolved images and spectra
with \chandra could clearly resolve this issue. Schulz et al. 2001 fully resolved the Orion Trapezium in the
X-ray band between 0.1 and 10 keV and found no diffuse emission between the Trapezium stars. Furthermore
four out of five of the brightest Trapezium stars showed hard X-ray spectra, with $\Theta^1$ Ori C being
the hottest star showing temperatures of up to 6$\times10^7$ K (paper~I). 

\begin{figure*}
\includegraphics[width=16.5cm]{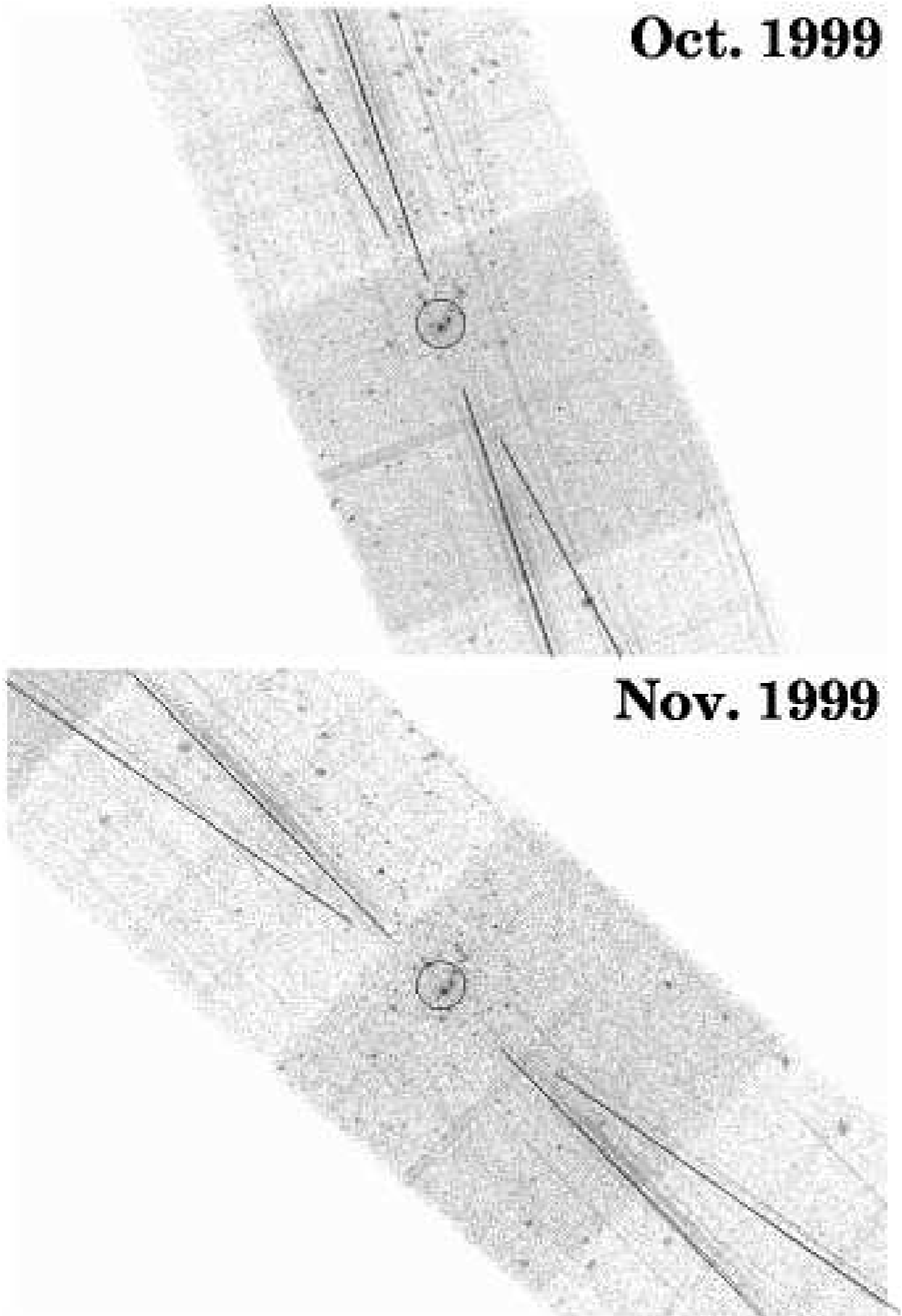}
\figcaption{Close-up view of the focal plane image near the zeroth order of the HETGS. The top
exposure shows the observation from October 1999, the bottom exposure the one from
November 1999, which was performed at a different roll angle. Sky north is up, east is left in
both images.}
\label{figure1}
\end{figure*}

In this paper we further investigate these issues through detailed modeling of High Energy Transmission Grating Spectrometer 
(HEGTS) spectra from the three X-ray brightest Trapezium stars $\Theta^1$ Ori A, C, and E (Schulz et al. 2001). 
These stars are excellent candidates for such a common study since they are presumably very young; the 
median age of the Orion Trapezium Cluster is 0.3 Myr (Hillenbrand 1997), they were born at the same time and 
should have had a quite similar initial chemical conditions. Their spectral types range from O6.5V
to B3.

\section{Chandra Observations}

We accumulate our spectra from two observations in the early phases of the \chandra mission. 
the first observation was performed on October 31st UT 05:47:21 1999 (OBSID 3) and lasted
50 ks. The second observation was obtained about three weeks later on November 24th UT 05:37:54 1999 (OBSID 4)
and lasted 33 ks.
For more details of the observations and some of the analysis threads we refer to paper~I
and Schulz et al. (2001).

\subsection{Data Analysis}

We re-processed the data using the most recently available calibration and CIAO\footnote{http://chandra.harvard.edu/CIAO2.3}
implementations
and produced  event lists containing the proper grating dispersion coordinates. The spectral extraction
was performed using CIAO tools and we also used custom software
for some of the broad band spectral analysis. The modeling of the spectra and its lines was done
using ISIS\footnote{http://space.mit.edu/ISIS}, the emission measure distribution was calculated
as in Huenemoerder, Canizares, $\&$ Schulz (2001). 
As already described in paper~I, the Orion
Trapezium is embedded in a cluster of fairly bright sources. We thus have to clean each spectrum from contributions
of interfering cluster sources, which would imitate lines by coincidence, as well as from dispersed photons of
other grating spectra crossing the dispersion track of interest. In our cases of interest the latter effect
was entirely eliminated by the energy discrimination by the CCDs. Figure 1 shows the HETGS focal plane
view near the zeroth order. In both exposures at the top and bottom part of the figure we encircled
the zeroth order positions of the main Trapezium stars. We also highlighted (for illustration purposes
only) the tracks of the dispersed spectra for the brightest source $\Theta^1$ Ori C. 

\vspace{0.3cm}
\noindent
\includegraphics[width=8.9cm]{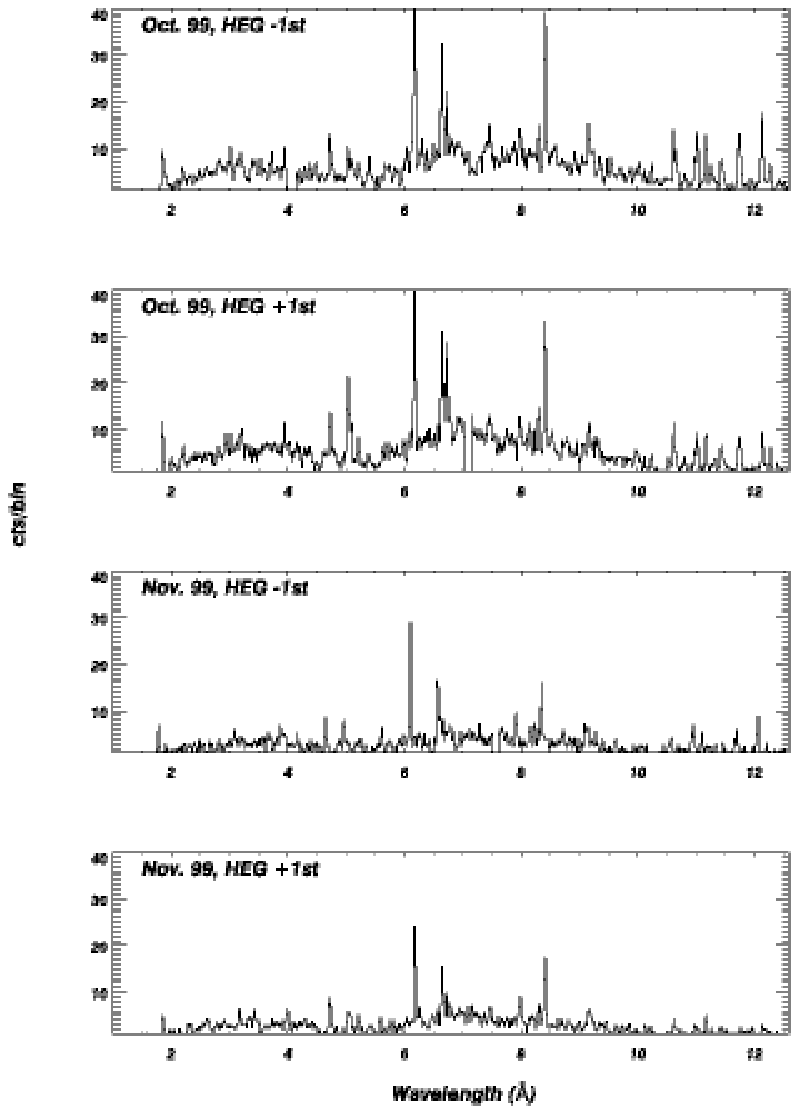}
\figcaption{HEG 1st order spectra of $\Theta^1$ Ori C. The top two panels show the +1st and -1st spectrum
of the first observation, the bottom two panels the same for the second observation.}
\label{figure2}
\vspace{0.3cm}
\noindent

\vspace{0.3cm}
\includegraphics[width=8.9cm]{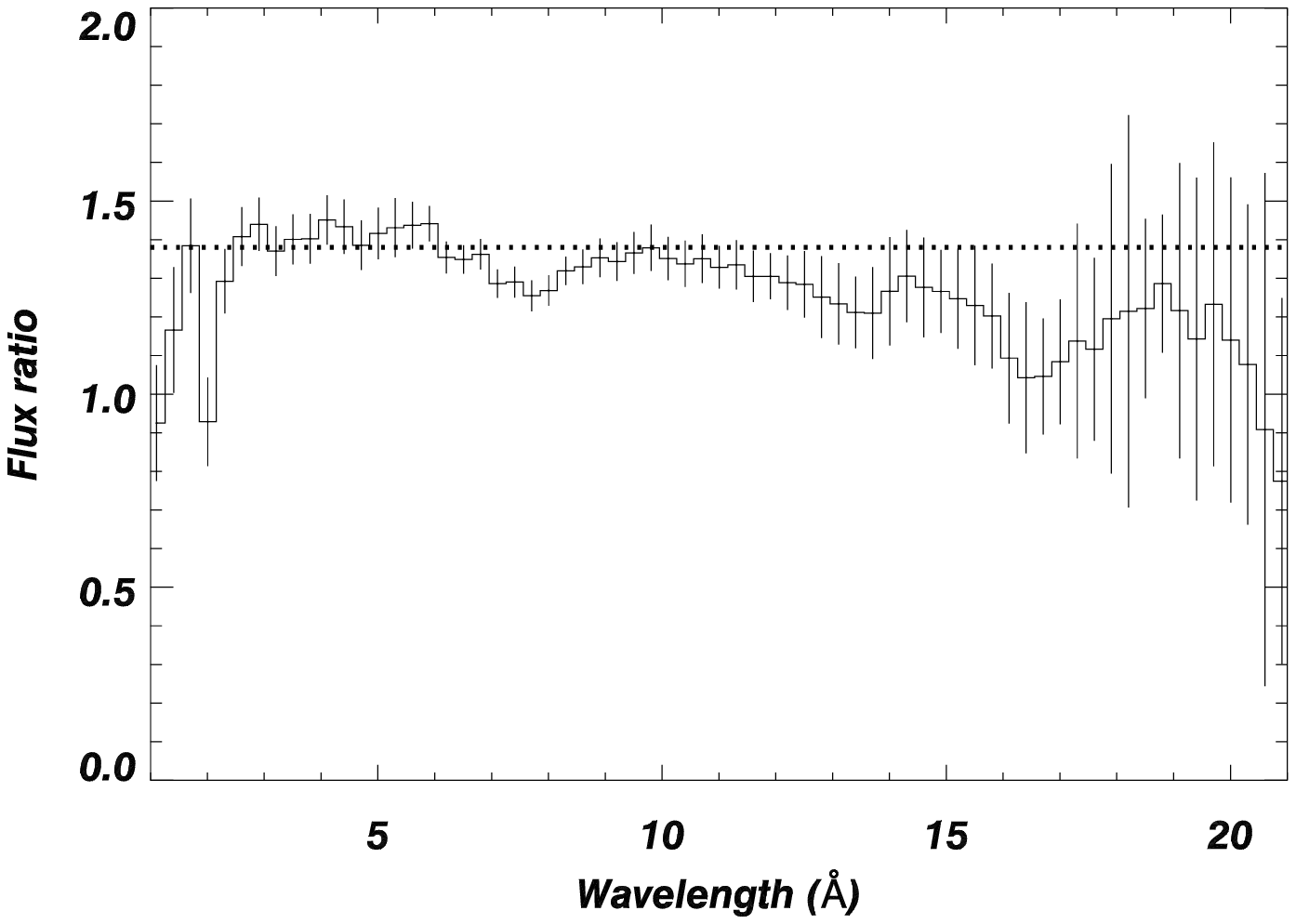}
\figcaption{The ratio of highly binned (0.5~\AA) and smoothed versions of the co-added spectra
from the first and second observation, which correspond to a wind-phase difference of approx. 0.45
in $\Theta^1$ Ori C.}
\label{figure3}
\vspace{0.3cm}

Many spectral tracks from various Trapezium Cluster stars are detected 
together with point sources at different off-axis positions which scatter throughout 
the entire array.
One of the differences to the extraction procedure outlined in paper~I is that we now have to 
also extract the spectra of $\Theta^1$ Ori A and E, which have a separation of a few arcsec only and 
are thus more likely to interfere with each other. We make extensive use  of the fact 
that we have observations at two different roll angles, 
which gives very good separation at 
least at one roll angle. Besides visual inspection, we also
use the fact that we know the relative fluxes of the two sources from the 0th order CCD spectra. 
In order to minimize the effect
of interfering point sources and parallel dispersion tracks of close sources we sometimes narrowed the 
cross-dispersion selection range. The standard selection range in cross-dispersion engulfs
95$\%$ of the flux and in cases where we narrowed this range we have to re-normalize the fluxes.
However we note that by correcting for the expected cross-dispersion profile, we add some 
systematic errors once we adjust final fluxes.  
We also correlated detected source point spread functions with the dispersion tracks of the three targets
and eliminated the data in the case of a true interference.  

\subsection{Raw and Exposure Corrected Spectra}

Figure~2 shows the first order HEG spectra of $\Theta^1$ Ori C for both observation periods after all cleaning
cycles. The weaker appearance of the November 1999 spectra compared to the October 1999 spectra is mostly
due the the difference in exposure, but for some lines there 
may also be variability related to the 15.4 day wind cycle. Here the October observation corresponds
to phase 0.82, the November observation to phase 0.37 using the ephemeris from Stahl et al. (1996).
In order to investigate this effect we added spectra from the MEG and HEG and divided the
exposure corrected flux spectra of the two phases. This spectral ratio is shown in Figure~3. 
We smoothed and binned the spectra into large (0.5~\AA~) bins.
The ratio shows that there is a flat part about 38\% above unity below 6~\AA~. 
The detailed analysis below shows that the changes are due to variable emissivity
at high temperatures.
The mean (rms) difference across the whole band is about 24$\%$. Schulz et al. (2001) 
report different fluxes in the zeroth of about 10$\%$, 
which is consistent given the uncertainties in that observation due to a over $20\%$ pile up fraction. 

\subsection{Line Widths}

A controversial issue in paper~I (see erratum in Schulz et al. 2003) was the analysis of the line widths 
in the case of $\Theta^1$ Ori C. Due to a software error the line width presented in that paper did 
not properly account
for the response of the instrument and, specifically between 10 and 12 ~\AA~ underestimated the 
amount of line blends. Thus the lines appeared wider than they really are.  Here we present the
analysis that properly includes the instrument response and and takes care of all the line 
blends.
The lines below about 13 ~\AA~ clearly appear unresolved.
Figure~4 (top) shows lines from three high energy H-like ions, from S XVI, Si XIV, and 
Mg XII. We chose these these lines because they are most likely not affected by blends.
The model used for the line fits is described in section 3.2.2 and it fits not only the local but the 
overall broad band continuum. Thus the model for all three lines comes from the same
fit (Figure~4 bottom). Shown are the co-added spectra and models (red curves). 
In addition we show a broad stretch
from the same model fit between 10.5 and 12 ~\AA~ showing many highly ionized and blended Fe states.
The model sufficiently describes these blends with no intrinsic broadening. 
For unresolved lines we can set a 90$\%$ confidence limit of the HWHM of 110 km s$^{-1}$ 
on average based on the statistical
properties of the spectrum.

\vspace{0.3cm}
\noindent
\includegraphics[width=8.9cm]{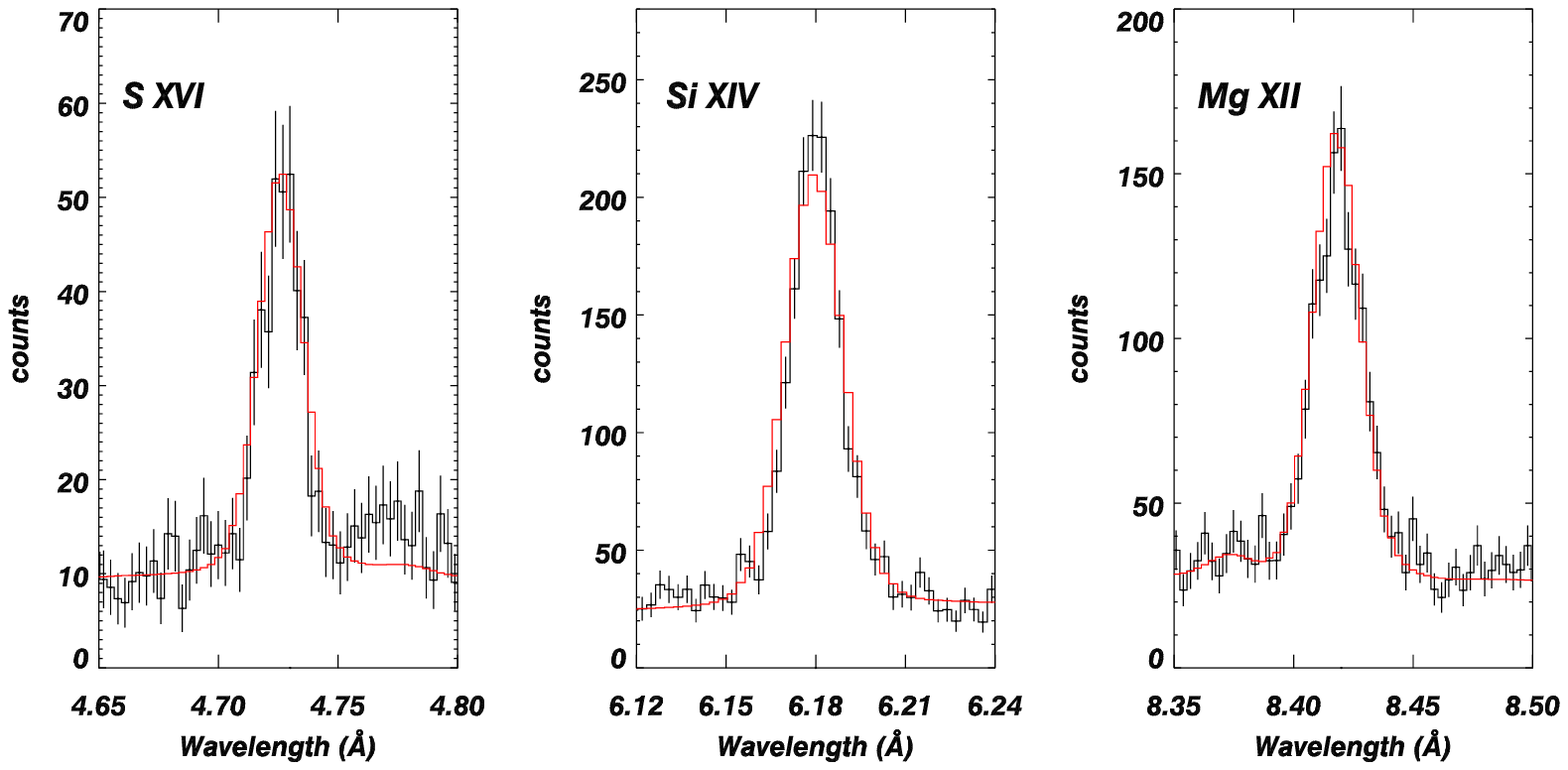}
\vspace{0.3cm}
\includegraphics[width=8.9cm]{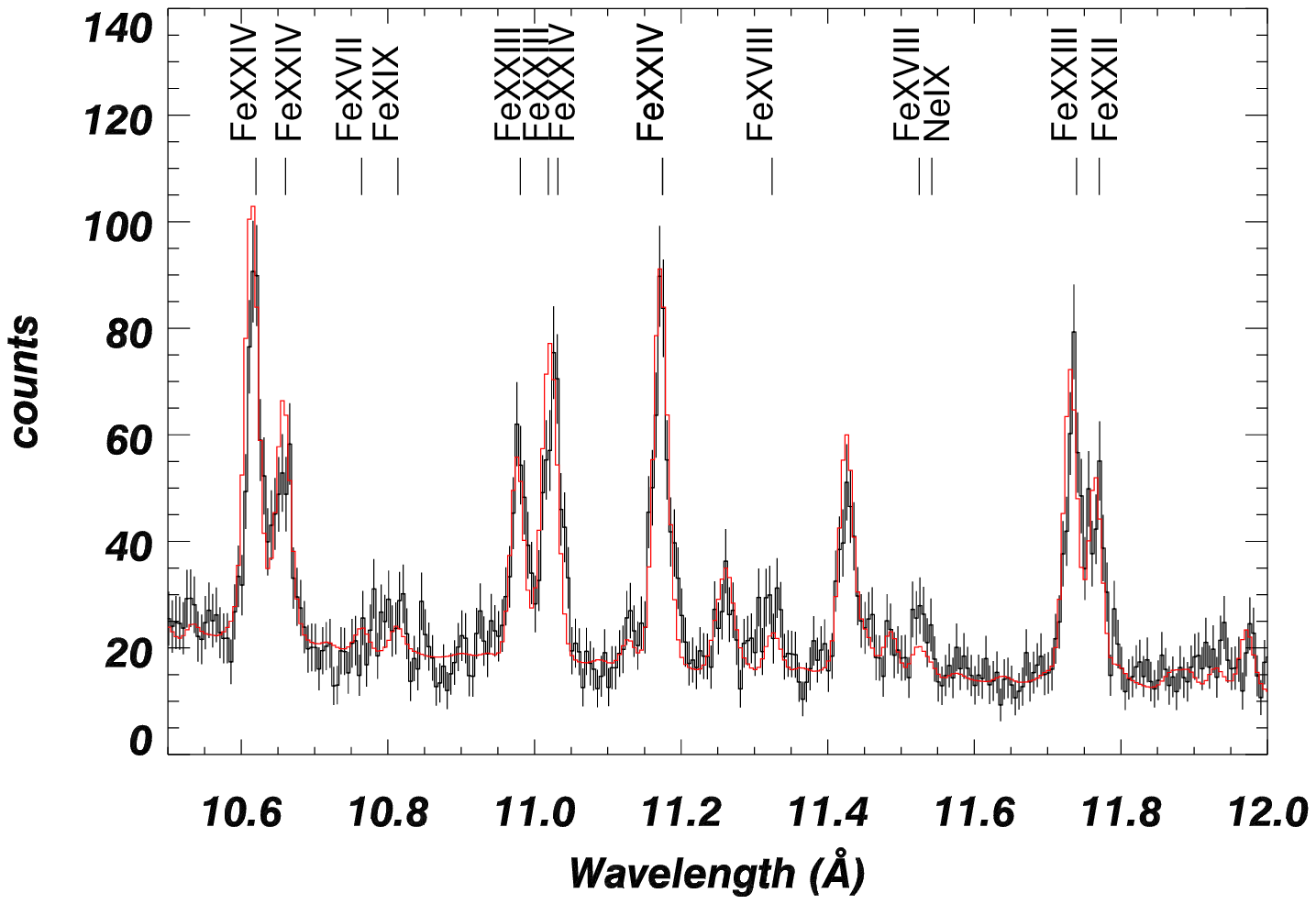}
\figcaption{{\bf Top:} Model fits of three unblended H-like lines. The red curve is the model
(as described in section 3.2.2). HEG and MEG data have been co-added, which allows to
oversample to profile with spectral bins of 0.003 ~\AA. {\bf Bottom:} The same
model and spectrum in the wavelength range between 10.5 and 12.0 ~\AA. It shows a critical stretch
of blended Fe lines. The model includes has zero broadening in the lines.
}
\label{figure4}
\vspace{0.3cm}

There is a second
population of lines at longer wavelengths which are resolved. The statistical quality
for many of them is marginal as in this bandpass the spectrum is effectively absorbed.
Figure~5 shows the two brightest resolved lines, one from an Fe XVII ion at 15.01~\AA~ and the 
O VIII line at 18.97 ~\AA. The Fe XVII can be fit with a HWHM of 460 km s$^{-1}$, the 
O VIII line with a HWHM of 850 km s$^{-1}$. The latter is also the only one we find
slightly blueshifted by about 240 km s$^{-1}$. In general it is expected to resolve lines better 
at higher wavelengths since the FWHM is constant with wavelength.
However these velocities exceed the confidence limit of 110 km s$^{-1}$ for the unresolved 
lines considerably and this indicates intrinsic broadening for some lower ionization states.   

\section{Spectral Analysis}

We divide the analysis of the $\Theta^1$ Ori C spectrum into two parts:  we model the 
broad band spectrum with various temperature components of hot plasmas, and we
construct an emission measure distribution from single line emissivities. For the latter we need a large number
of significant line fluxes, which we only have available in the case of $\Theta^1$ Ori C. For the other hot
stars we then compute variations of the broad band plasma model derived from the $\Theta^1$ Ori C spectrum.      

\includegraphics[width=8.9cm]{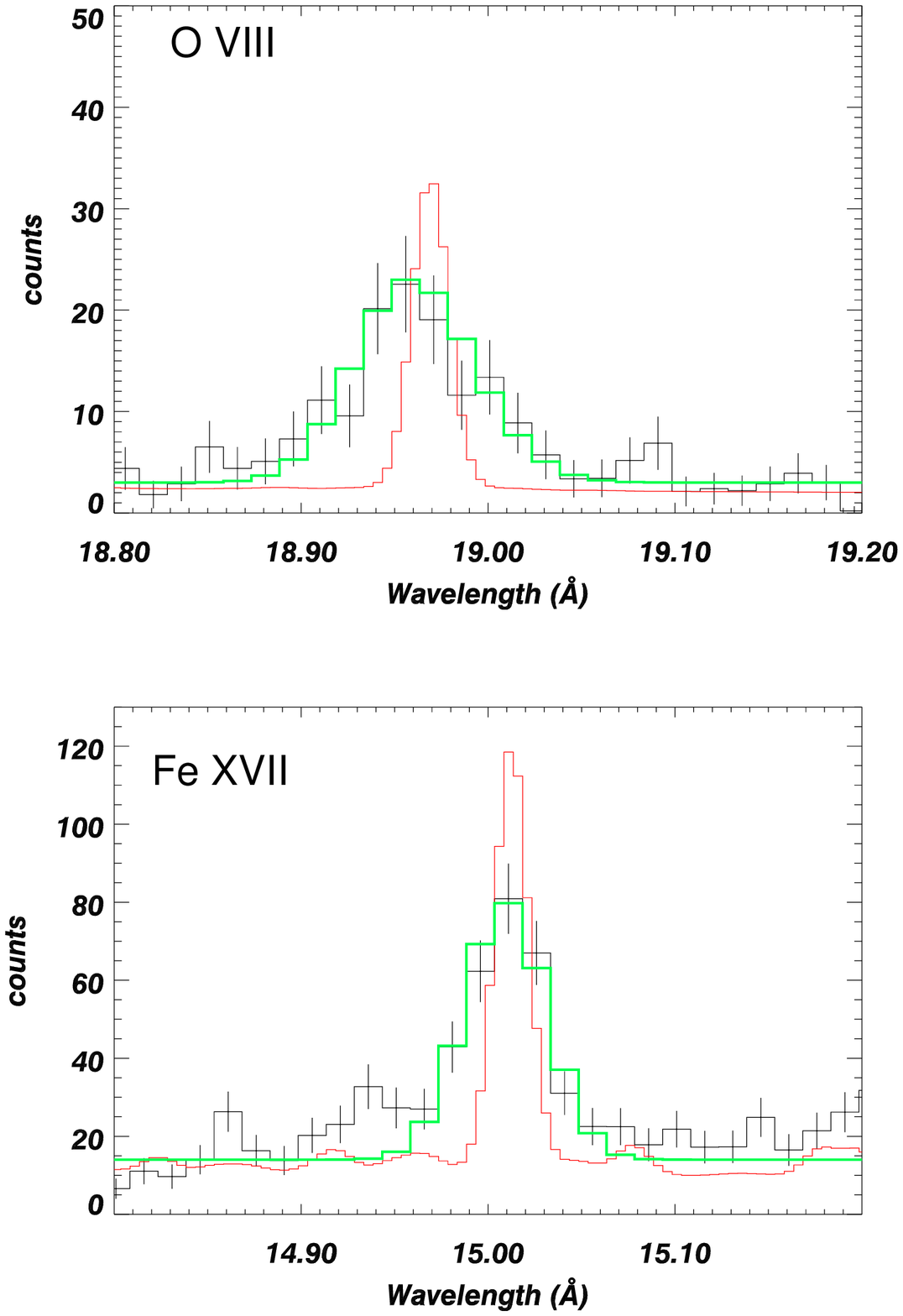}
\figcaption{The two most significant soft lines in the spectrum. The red line shows the model
used in Figure 4 with zero line broadening. The green line is a local Gaussian fit to the data.
A comparison shows that both lines appear resolved and show significant broadening. The O VIII
in the top diagram also appears blue-shifted.
}
\label{figure5}

\subsection{The Spectral Model}

For the spectral modeling we exclusively use the Astrophysical Plasma Emission Code and Database 
(APEC and APED\footnote{http://hea-www.harvard.edu/APEC}) described by Smith et al. (2001). The database
is available in ISIS and we can compute emissivities for a collisionally ionized equillibrium plasma in terms 
of temperature, density, and various abundance distributions. In paper~I we deduced that   
these spectra require a range of temperatures. For the model we assume that  
all emitting plasmas contributing to the spectrum are in collisional ionization equillibrium. This
means that a thermal plasma is in a stable ionization state under the coronal approximation. In this 
approximation it is assumed that the dominant processes are collisional excitation or ionization from
the ground state balanced by radiative decay and recombination. Photo-ionization and -excitation
as well as collisional ionization from excited states are assumed to be negligible.   
Note that due to the usually quite strong UV radiation field near the stellar surface this assumption
is not quite true and the meta stable forbidden lines in the He-like triplets may be affected. We will
discuss the He-like triplets in a separate section.

From over 80 significant 
lines we calculate a differential emission measure (DEM) distribution for the two phases 
separately and for the phase averaged spectra. In the other two stars,
$\Theta^1$ Ori A and E, we lack this large number of lines and cannot calculate such a significant DEM. 
We therefore approximate the phase averaged DEM of $\Theta^1$ Ori C by a few  
temperature components for which we compute model spectra.
We allow as many temperature components as 
necessary to account for 95$\%$ of the lines (3 $\sigma$) observed. 
Once adding temperature components does not improve the fit, 
we allow the abundance distribution to vary against a solar element abundance distribution. 
A similar procedure was followed to also fit the spectra obtained for $\Theta^1$ Ori A and E. 

\vspace{0.3cm}
\includegraphics[width=8.1cm]{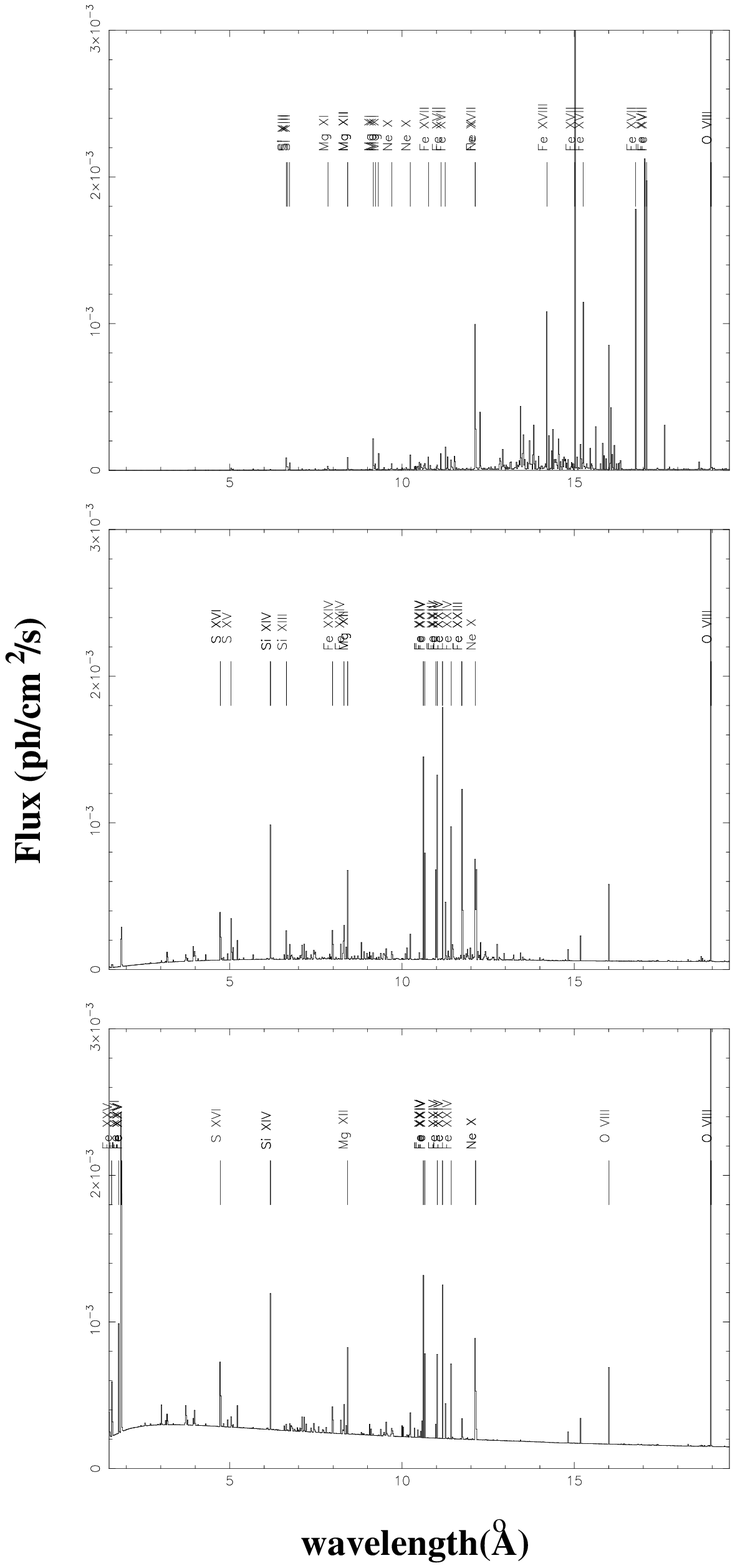}
\figcaption{APED model spectra at different temperatures and with solar abundances. The top spectrum was calcula
ted at
6 10$^6$ K, the middel spectrum corresponds to 1.5 10$^7$ K, and the bottom spectrum to 6.0 10$^7$ K.
The line width was fixed for Doppler velocities of 200 km s$^{-1}$.}
\label{figure6}
\vspace{0.3cm}

\vspace{0.3cm}
\includegraphics[width=8.5cm]{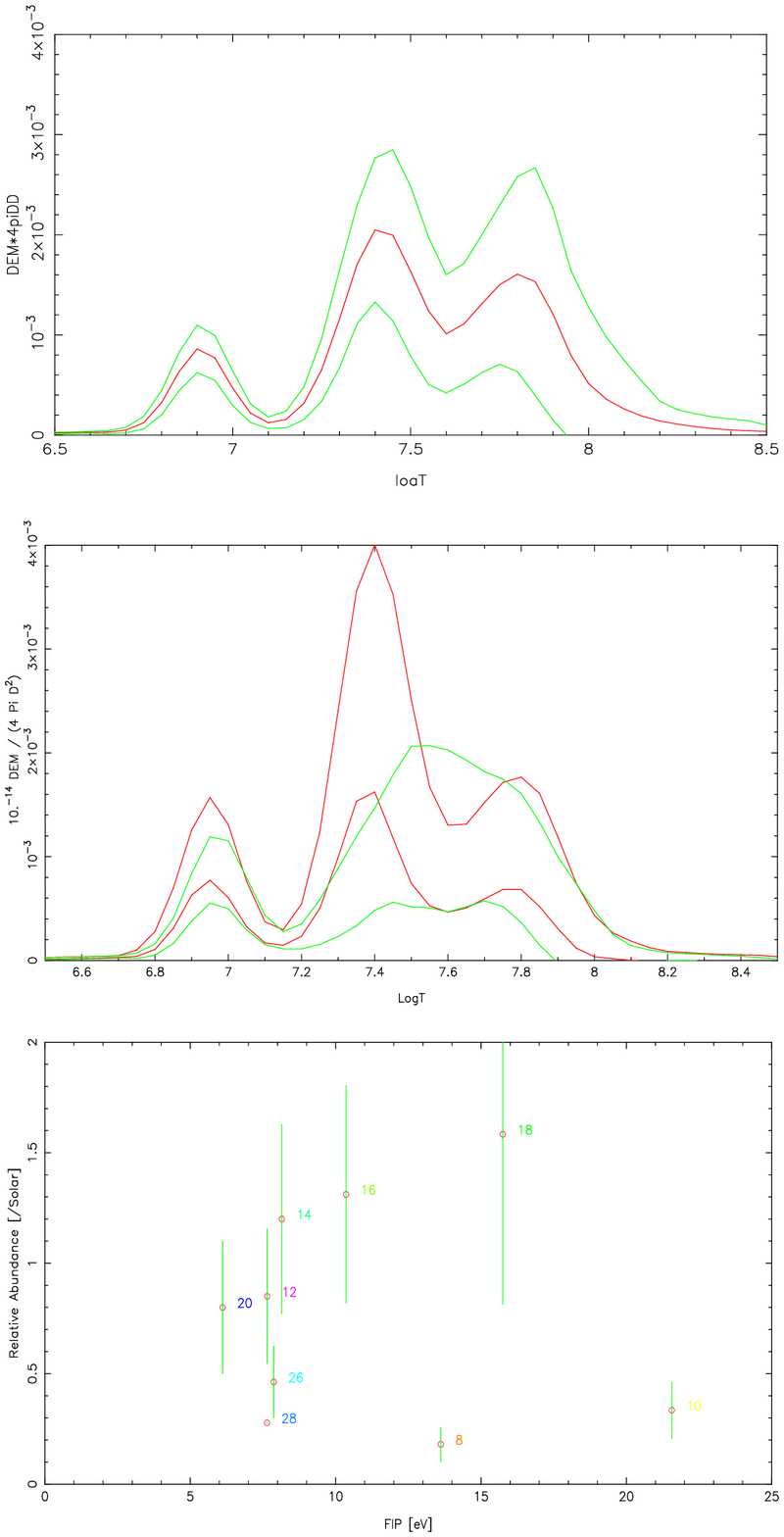}
\figcaption{{\bf Top:} The distribution of DEMs versus the log of the temperature.
The middle curve represents the mean distribution, the top and bottom curve the
one $\sigma$ uncertainty envelope from MonteCarlo fitting. {\bf Middle:} The solid lines represent the upper
and
lower error envelopes of the DEM from the phase 0.38 spectrum, the dotted lines
the ones for the phase 0.76 spectrum. {\bf  Bottom:}
The distribution of abundances normalized to solar (which represents 1) versus the
ionization potential for various abundant elements for the phase averaged DEM.}
\label{figure7}

Figure~6 demonstrates three typical cases of spectra within 
the temperature range we found in paper~I.
All three cases are computed in the low density limit, i.e. for densities below  1$\times10^{10}$ cm$^{-3}$. 
The spectrum in the top panel has a low temperature of
6 MK. It hardly shows thermal continuum flux, but many lines - mostly from transitions of Fe XVII -
in the region around 15~\AA~. The middle panel shows an intermediate temperature spectrum of 
15 MK. We now do not observe Fe transitions below Fe XXII, but very strong Fe XXIV (Li-like)
lines in the range between 8 and 12~\AA~ as well as a faint Fe XXV (He-like) line at 1.85 ~\AA.
We also see stronger H- and He-like lines from Si, S, Ar and even Ca. 
There is already a significant
fraction of a thermal continuum. At 60 MK we observe weaker Fe XXIV lines in the 10 to 12~\AA~ region,
but very strong Fe XXV and still strong lines from other high Z elements. Most important is the  
existence of a strong continuum. This continuum as well as its high energy cut-off value increases 
with temperature. The value of this cutoff in 
connection with the Fe XXV emissivity is a sensitive benchmark for the high temperature 
of the plasma. The continuum also allows measurements of abundances relative to H, since it
is dominated by thermal bremsstrahlung.
O VIII (Lyman $\alpha$ at 19 ~\AA) is contributing at all temperatures, at higher temperatures the 
Lyman $\beta$ and $\gamma$ lines become strong as well.

\subsection{$\Theta^1$ Ori C}

We assumed a line-of-sight column
density of 1.93$\times10^{21}$ cm$^{-2}$ throughout the entire observation.
From the measured color excess of E$_{B-V}$ = 0.32 (Bolin$\&$Savage 1981) and using
the correlation between N$_H$ and E$_{B-V}$ derived from L$_{\alpha}$ measurements in hot stars
by Savage $\&$ Jenkins 1972 we expect 1.7$\times10^{21}$ cm$^{-2}$; from Spitzer (1978) we expect
1.9$\times10^{21}$ cm$^{-2}$; from Predehl $\&$ Schmitt (1995) we expect 1.4$\times10^{21}$ cm$^{-2}$ for
the average interstellar medium (R$_v$ = 3.1) and 2.4$\times10^{21}$ cm$^{-2}$ for dense molecular
clouds (R$_v$ = 5.1). From these numbers an additional opacity in the star cannot
be concluded. 

\subsubsection{Emission Measure Distributions}

The
procedure for the DEM derivation from an X-ray spectrum is described in Huenemoerder, Canizares,
$\&$ Schulz (2002). The volume emission measure (VEM) is the volume integral over the product of
electron (n$_e$) and hydrogen (n$_h$) densities at a given temperature 
(note: since the emissivity of most emission
lines are only weakly dependent on n$_e$ in the expected temperature and density regime, 
we can as well ignore it). We define a 
temperature grid of 60 points spaced by 0.05 in log T. The DEM is the derivative
of the volume emission measure with respect to log T. It incorporates the emission measure and abundance
of many ions and reflects the energy balance of the plasma with respect to temperature.

In Table 1 we show the list of detected
lines with measured and predicted line positions and fluxes for the combined spectra.
We also, using APED, computed a
temperature of maximum emissivity from each ion. The tables are
sorted by ion species, which approximately reflects temperature as well.

The first part of Table I shows lines from all Fe ions accepted by the fit with
one $\sigma$  confidence errors for the line fluxes. It is specifically remarkable to see a
large number of ion states larger than Fe XXII, specifically Fe XXIV, and ion states
lower than XIX, specifically Fe XVII.
The fractional abundances (i.e. the ratio of integrated flux of a specific ion species over
all Fe emission) show a dominance
of Fe XXIII to Fe XXV ions ( 0.36) as well as Fe XVII to Fe XIX (0.58), but a significant lack
of Fe XX to XXII ions (0.06).
The line positions agree well within the expected
uncertainties with the positions produced
by APED. The second part of Table I shows the list of ions with Z lower than 26.
The results are very
similar to what we observe for Fe ions, i.e. lines from low and high temperatures dominate
the fractional abundances.

Figure 7 (top) shows the DEM versus temperature over a range
from 3 to 300 MK. The middle (red)  line shows the distribution, the upper and lower lines (green)
trace the 90$\%$ uncertainty range. There are clearly three peaks visible at 7.9$\pm$0.2 MK,
26.3$\pm$1.8 MK, and 66.1$\pm$11.2 MK. The emissivity is dominated by plasmas at temperatures
higher than 15 MK with a temperature tail that allows the distribution to exceed 100 MK.
There is a prominent gap at 7.15 MK that divides the DEM into
emission from high and low temperature plasmas. This drop roughly corresponds to the
missing emissivities from Fe XX to XXII ions in Table 1. The low temperature peak
incorporates most ions from O to Si and  Fe XVII to Fe XXII, while the high
temperature DEM corresponds to Fe XXIII to Fe XXV, Si to Ca.
There is also an incision
in the high temperature part of the DEM that roughly corresponds to the weak showing
of S XVI and Ca XIX in one of the spectra (see below).

The middle part of Figure 7 show the DEMs for two phases separately. The solid lines
show the upper and lower error limits of the DEM for phase 0.82, the
dotted lines the same for phase 0.37. There are a few remarkable characteristics.
The incision between high and low
temperature at 7.15 MK persists in both phases and the low temperature
emissivity is similar. The
high temperature emissivity shows a significant difference in that the middle peak is much more
pronounced in phase 0.82 whereas the high temperature peaks appear unchanged.
This means the overall variability in line emissivity is predominately in the
Fe XXII and Fe XXIV lines as well as in the Si XIV, where we observe the bulk of the emissivity.

The DEM fit also adjusts abundances (bottom of Figure 7) with
respect to solar abundances. Here we find for
Mg, Si, S, Ar, Ca ions no significant deviation from solar values. For O we find
a factor of 0.2$\pm$0.1, for Ne and Fe a factor of 0.5$\pm$0.1 underabundance.
We find this distribution in both observations. Additional uncertainties
in these values may stem from uncertainties in the ionization balance (Mazzotta et al. 1998).
However we also have to stress the point that once we apply the spectra for each ion
fitted in the DEM analysis we find a sufficient fit to the total spectrum (see also below).

The determination of the continuum in the DEM analysis is an iterative process assuming
that all continuum contributions come from the same thermal plasma generating the emission lines.
Contributions from a non-thermal component critically affect the Fe abundances. A predicted
shape of $\propto E^{-1/2}$ of a Compton component (Chen and White 1991) most prominently
contributes to the continuum near the Fe XXV line. However the continuum level there is
already quite determined by the fits to the Fe XXIV lines and their adjacent continuum
levels. We estimate that the contribution cannot be more than about 1$\%$ to the total X-ray flux.

\subsubsection{Phase-averaged Plasma Model}

Figure 8 shows the phase averaged count spectrum binned by
0.005~\AA. MEG and HEG have been added.
In this approach we model this spectrum with a few constrained components by approximating the
emission measure above. This section in this respect does not produce much new information, but
verifies the method used for the fainter stars. However, for reasons of consistency we use the luminosities
and fluxes determined in this section for further discussion.
We rebin the above DEM into a few coarse intervals and calculate
a model component for each DEM interval. We could calculate the model
directly from each DEM bin, i.e. from the log T (K) = 0.05 grid. However the idea is to
produce a simplified model in good approximation to the DEM.
The models were again calculated using the APED database, folded through the
spectral response function and were then fit to the measured spectra.

\vbox{
\begin{center}
{\sc TABLE~1 FE IONS DETECTED IN $\Theta^1$ Ori C}
\vskip 4pt
\begin{tabular}{lcccc}
\hline
\hline
 ion & $<$log T$>$& $\lambda_o$ & $\lambda_{maes}$  & flux \\
     & K & \AA  & \AA & 10$^{-5}$ ph s$^{-1}$~cm$^{-2}$ \\
\hline
 & & & & \\
Fe XXV &  7.8 & 1.860 &   1.858 &   7.534$\pm$  1.225 \\
Fe XXIV & 7.4 & 7.169 &   7.170 &   1.341$\pm$  0.252 \\
Fe XXIV & 7.4 & 7.996 &   7.989 &   1.774$\pm$  0.232 \\
Fe XXIV & 7.4 & 8.316 &   8.233 &   0.884$\pm$  0.200 \\
Fe XXIV & 7.4 & 8.233 &   8.285 &   0.248$\pm$  0.177 \\
Fe XXIV & 7.4 & 8.285 &   8.304 &   0.924$\pm$  0.348 \\
Fe XXIV & 7.4 & 10.619 &  10.620 &   5.825$\pm$  0.585 \\
Fe XXIV & 7.4 & 10.663 &  10.660 &   3.453$\pm$  0.494 \\
Fe XXIV & 7.4 & 11.029 &  11.032 &   3.359$\pm$  1.209 \\
Fe XXIV & 7.4 & 11.176 &  11.175 &   7.026$\pm$  0.718 \\
Fe XXIII & 7.2 & 8.304 &   8.316 &   1.484$\pm$  0.364 \\
Fe XXIII & 7.2 & 8.815 &   8.816 &   1.139$\pm$  0.243 \\
Fe XXIII & 7.2 & 10.981 &  10.981 &   3.029$\pm$  0.456 \\
Fe XXIII & 7.2 & 11.019 &  11.019 &   2.592$\pm$  0.000 \\
Fe XXIII & 7.2 & 11.736 &  11.739 &   5.927$\pm$  0.693 \\
Fe XXIII & 7.2 &12.161 &  12.160 &   3.388$\pm$  0.950 \\
Fe XXII & 7.1 & 8.975 &   8.974 &   0.561$\pm$  0.224 \\
Fe XXII & 7.1 & 11.770 &  11.771 &   4.141$\pm$  0.602 \\
Fe XXI & 7.1 & 12.284 &  12.286 &   4.355$\pm$  0.908 \\
Fe XX &  7.0 & 9.219 &   9.194 &   1.151$\pm$  0.272 \\
Fe XX & 7.0 & 14.267 &  14.248 &   0.614$\pm$  0.667 \\
Fe XIX & 6.9 & 10.816 &  10.814 &   0.862$\pm$  0.307 \\
Fe XIX & 6.9 & 13.462 &  13.460 &   3.366$\pm$  0.970 \\
Fe XIX & 6.9 & 13.497 &  13.490 &   3.536$\pm$  0.950 \\
Fe XIX & 6.9 & 13.518 &  13.520 &   5.639$\pm$  1.157 \\
Fe XIX & 6.9 & 15.079 &  15.086 &   2.156$\pm$  1.182 \\
Fe XIX & 6.9 & 16.110 &  16.098 &   1.637$\pm$  1.006 \\
Fe XVIII & 6.8 & 11.326 &  11.324 &   1.433$\pm$  0.372 \\
Fe XVIII & 6.8 & 11.527 &  11.525 &   1.532$\pm$  0.521 \\
Fe XVIII & 6.8 & 14.208 &  14.208 &   3.472$\pm$  1.057 \\
Fe XVIII & 6.8 & 14.256 &  14.268 &   0.943$\pm$  0.626 \\
Fe XVIII & 6.8 & 14.534 &  14.535 &   1.493$\pm$  0.658 \\
Fe XVIII & 6.8 & 16.071 &  16.071 &   4.501$\pm$  1.445 \\
Fe XVIII & 6.8 & 16.159 &  16.184 &   1.100$\pm$  0.804 \\
Fe XVII & 6.8 &10.770 &  10.764 &   0.565$\pm$  0.289 \\
Fe XVII & 6.7 & 12.124 &  12.129 &   8.754$\pm$  1.168 \\
Fe XVII & 6.7 & 12.266 &  12.265 &   2.184$\pm$  0.580 \\
Fe XVII & 6.7 & 15.014 &  15.014 &  18.463$\pm$  2.347 \\
Fe XVII & 6.7 & 15.261 &  15.261 &   5.316$\pm$  1.181 \\
Fe XVII & 6.7 & 16.780 &  16.775 &  10.039$\pm$  2.255 \\
Fe XVII & 6.7 & 17.051 &  17.051 &   8.044$\pm$  2.099 \\
Fe XVII & 6.7 & 17.096 &  17.095 &   6.147$\pm$  1.921 \\
 & & & & \\
\hline
\end{tabular}
\end{center}
\normalsize
}

\vbox{
\begin{center}
{\sc TABLE~1 CONT.: LOWER ($< 26$) Z  IONS}
\vskip 4pt
\begin{tabular}{lcccc}
\hline
\hline
 ion & $<$log T$>$& $\lambda_o$ & $\lambda_{maes}$  & flux \\
     & K & \AA  & \AA & 10$^{-5}$ ph s$^{-1}$~cm$^{-2}$ \\
\hline
 & & & & \\
Ca XX & 7.8 &  3.024 &   3.024 &   0.602$\pm$  0.285 \\
Ca XIX & 7.5 & 3.177 &   3.179 &   1.206$\pm$  0.326 \\
Ca XIX & 7.5 & 3.207 &   3.209 &   0.752$\pm$  0.292 \\
Ar XVII & 7.4 &  3.365 &   3.372 &   0.303$\pm$  0.325 \\
Ar XVIII & 7.7 & 3.731 &   3.735 &   1.002$\pm$  0.359 \\
Ar XVII & 7.4 & 3.949 &   3.943 &   1.695$\pm$  0.380 \\
Ar XVII & 7.3 & 3.966 &   3.960 &   1.243$\pm$  0.489 \\
S XVI &  7.6 & 4.733 &   4.729 &   5.327$\pm$  0.619 \\
S XV &  7.2 & 5.039 &   5.039 &   5.201$\pm$  0.710 \\
S XV &  7.2 & 5.067 &   5.065 &   1.798$\pm$  0.486 \\
S XV &  7.2 & 5.102 &   5.100 &   2.378$\pm$  0.541 \\
Si XIV & 7.4 &  5.218 &   5.217 &   2.394$\pm$  0.510 \\
Si XIV &  7.4 & 6.180 &   6.181 &  11.662$\pm$  0.562 \\
Si XIII & 7.0 & 5.681 &   5.684 &   0.836$\pm$ 0.276 \\
Si XIII & 7.0 &  6.648 &   6.648 &   7.432$\pm$  0.472 \\
Si XIII & 7.0 &  6.687 &   6.686 &   2.515$\pm$  0.328 \\
Si XIII & 7.0 &  6.740 &   6.740 &   3.717$\pm$  0.342 \\
Al XII & 7.0 &  7.757 &   7.760 &   0.595$\pm$  0.196 \\
Al XII & 6.9 &  7.872 &   7.874 &   0.256$\pm$  0.184 \\
Mg XII & 7.2 &  7.106 &   7.105 &   1.554$\pm$  0.235 \\
Mg XI & 6.9 &  7.850 &   7.850 &   0.747$\pm$  0.206 \\
Mg XII & 7.2 &  8.419 &   8.420 &   8.406$\pm$  0.484 \\
Mg XI & 6.8 &  9.169 &   9.168 &   3.085$\pm$  0.370 \\
Mg XI & 6.8 &  9.231 &   9.230 &   1.966$\pm$  0.314 \\
Mg XI & 6.8 & 9.314 &   9.310 &   0.917$\pm$  0.305 \\
Ne X & 7.0 &  9.481 &   9.478 &   1.351$\pm$  0.271 \\
Ne X & 7.0 &  9.708 &   9.710 &   0.889$\pm$  0.250 \\
Ne X & 7.0 & 10.240 &  10.240 &   1.822$\pm$  0.334 \\
Ne IX & 6.6 & 11.544 &  11.542 &   0.595$\pm$  0.454 \\
Ne X & 6.9 & 12.132 &  12.144 &   5.325$\pm$  1.131 \\
Ne IX & 6.6 & 13.447 &  13.435 &   3.804$\pm$  0.995 \\
Ne IX & 6.6 & 13.550 &  13.550 &   3.827$\pm$  0.962 \\
O VIII & 6.7 & 14.821 &  14.840 &   0.930$\pm$  0.622 \\
O VIII & 6.7 & 15.176 &  15.195 &   2.767$\pm$  0.910 \\
O VIII & 6.7 & 16.006 &  16.010 &   3.934$\pm$  1.200 \\
O VII & 6.4 & 18.627 &  18.635 &   4.067$\pm$  2.007 \\
O VIII & 6.7 & 18.973 &  18.970 &  21.143$\pm$  3.008 \\
O VII & 6.3 & 21.602 &  21.600 &  10.230$\pm$  5.583 \\
O VII & 6.3 & 21.804 &  21.799 &  10.451$\pm$  5.991 \\
 & & & & \\
\hline
\end{tabular}
\end{center}
\normalsize
}

\begin{figure*}
\includegraphics[width=15cm]{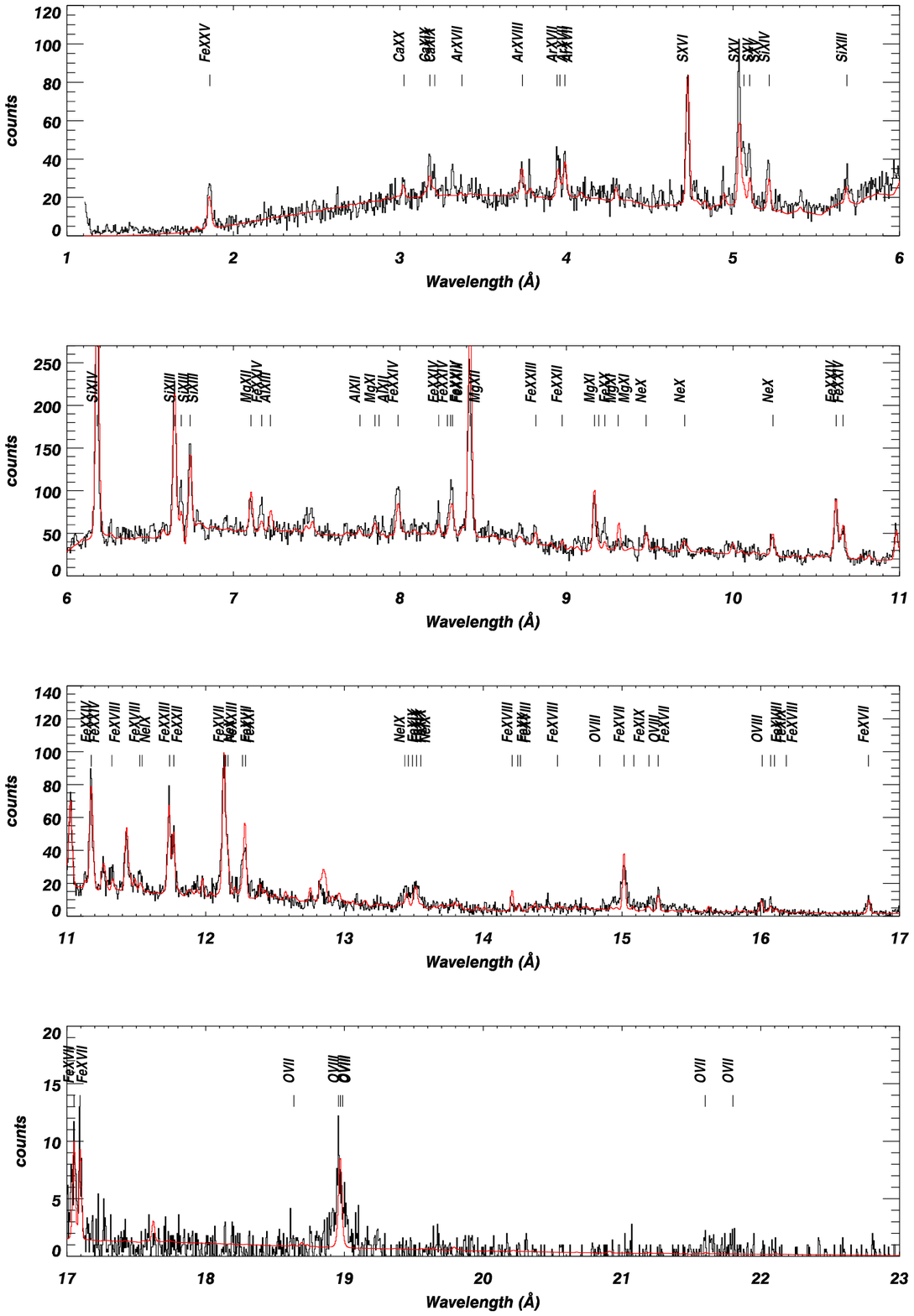}
\figcaption{The measured phase averaged count spectrum over the full exposure for $\Theta^1$ Ori C. The bins
of the spectrum have a value of 0.005~\AA. The red line is the approximated phase-averaged plasma model
deduced for this spectrum.}
\label{figure8}
\end{figure*}

\vspace{0.3cm}
\vbox{
\footnotesize
{\sc TABLE~2 TEMPERATURE COMPONENTS FROM MODEL FITS}
\vskip 4pt
\begin{tabular}{lcccc}
\tableline
\tableline
star  &comp. & log T         &  norm                   & L$_x$ (1-10 keV)\\
      &$\#$  &  K  & erg s$^{-1}$~cm$^{-2}$   &\ergsec \\
\tableline
$\Theta^1$ Ori C & 1 & 6.79$\pm$0.06 & 6.55$\times10^{-13}$ & 7.93$\times10^{30}$ \\
                 & 2 & 6.97$\pm$0.07 & 2.22$\times10^{-12}$ & 2.69$\times10^{31}$ \\
                 & 3 & 7.22$\pm$0.07 & 1.24$\times10^{-12}$ & 1.50$\times10^{31}$ \\
                 & 4 & 7.47$\pm$0.08 & 4.62$\times10^{-12}$ & 5.60$\times10^{31}$ \\
                 & 5 & 7.64$\pm$0.06 & 2.51$\times10^{-12}$ & 3.04$\times10^{31}$ \\
                 & 6 & 7.82$\pm$0.07 & 5.20$\times10^{-12}$ & 6.30$\times10^{31}$ \\
\tableline
$\Theta^1$ Ori E & 1 & 6.63$\pm$0.08 & 1.55$\times10^{-13}$ & 1.87$\times10^{30}$ \\
                 & 2 & 7.02$\pm$0.08 & 3.78$\times10^{-13}$ & 4.60$\times10^{30}$ \\
                 & 3 & 7.31$\pm$0.07 & 4.88$\times10^{-13}$ & 5.92$\times10^{30}$ \\
                 & 4 & 7.49$\pm$0.06 & 1.85$\times10^{-13}$ & 2.23$\times10^{30}$ \\
                 & 5 & 7.67$\pm$0.06 & 1.31$\times10^{-12}$ & 1.59$\times10^{31}$ \\
\tableline
$\Theta^1$ Ori A & 1 & 6.68$\pm$0.09 & 3.41$\times10^{-14}$ & 4.13$\times10^{29}$ \\
                 & 2 & 7.06$\pm$0.08 & 3.54$\times10^{-13}$ & 4.29$\times10^{30}$ \\
                 & 3 & 7.34$\pm$0.08 & 1.08$\times10^{-13}$ & 1.31$\times10^{30}$ \\
                 & 4 & 7.48$\pm$0.07 & 4.28$\times10^{-13}$ & 5.18$\times10^{30}$ \\
                 & 5 & 7.63$\pm$0.07 & 2.76$\times10^{-13}$ & 3.35$\times10^{30}$ \\
\tableline
\end{tabular}
\normalsize
}
\vspace{0.3cm}

We fit only one 
component at a time (keeping the parameters of the other ones fixed) and repeated this step
with the other components until the final spectrum meets the criteria below. 
To sensitively constrain the model we use the DEMs above, which
helps us to constrain the relative strengths of the components: the level of the continuum,
the position of its high-energy cut-off, the strength of the Fe XXV lines and various temperature
dependent line ratios. 

Our model reproduces all major properties of the spectrum. There are several 
small deviations relating to the fact that the model cannot yet reproduce all subtle details.  
The results are summarized in Table 2. The stated temperature uncertainties are
large which is a consequence of the large error envelopes of the DEM in combination with 
a now very coarse temperature grid. The stated fluxes and luminosities
have errors between 5 and 15$\%$. 
Several key properties in the spectrum are well modeled: the total flux derived from the model and 
from the exposure corrected data in the range of 1 to 25~\AA~ agree within 2$\%$, the continuum 
is well represented and the line ratios agree within 10$\%$.

The final result is overplotted in Figure 8.
The thermal flux continuum shows a cut-off around 3.4~\AA. In order to get this cut-off we 
need a high temperature of 66.5 MK. The shallow decline of the continuum below the cut-off 
requires stronger intermediate temperature components (43.8 and 29.5 MK), which are constrained by
by line ratios of Si, S, and Fe lines 
as well as the fact that the sum of the Fe XXV line fluxes of these two high temperature components 
need to match the observed line flux. These three components account for almost all of the 
continuum providing 75$\%$ of the star's X-ray luminosity. 
The low temperature components
at 6.1, 9.3 and 16.5 MK produce many lines between 9 and 20~\AA~ but account only
for a fraction of the total luminosity. In fact, if the star did not have hot components,
it would be faint. From Table 2 the X-ray luminosity accounting for the low temperature DEM
component only is 3.5$\times10^{31}$ \ergsec. Using the bolometric luminosity as listed
by Bergh\"ofer, Schmitt, and Cassinelli (1996) we find -7.2 for the low temperature log $L_x/L_{bol}$. 
More than 85$\%$ of the
luminosity is radiated by plasmas with temperatures higher than 1.5$\times10^7$ K, and 75$\%$ from
temperatures higher than 3$\times10^7$ K. This is consistent with the results from the
zero order CCD spectra (Schulz et al. 2001). 
The abundances had to be adjusted during the fit as well and the result is similar to the 
distribution observed in the DEM analysis.

\includegraphics[width=8.5cm]{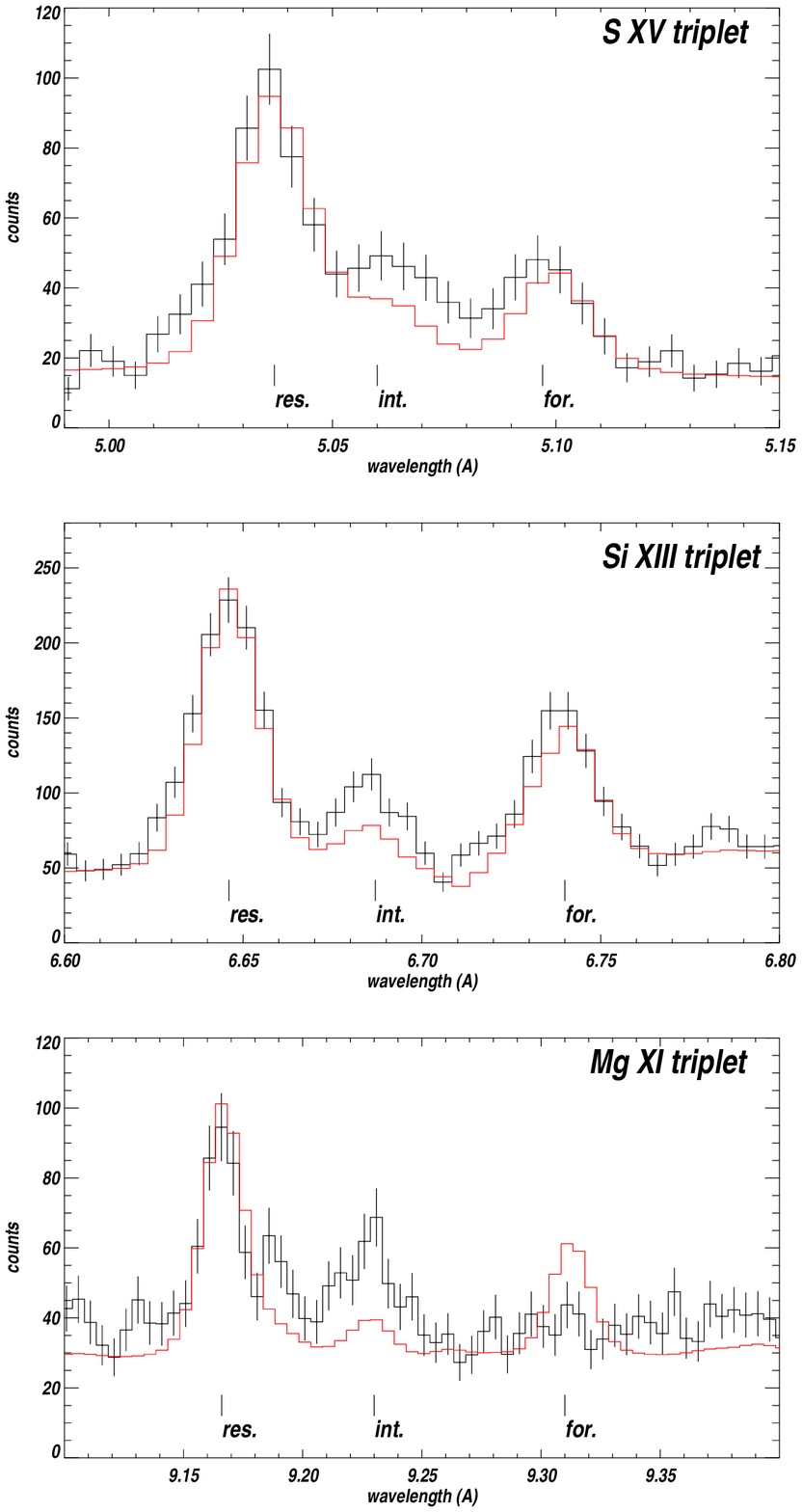}
\figcaption{The He-like triplets from S (top), Si (middle), and Mg (bottom) from the spectrum in figure 8.
The model here has been optimized such for each triplet that it fits the recombination line.}
\label{figure9}

\subsubsection{Optical depths, Formation Radii and Densities}

Line ratios may be used
to place limits on optical depths, line formation radii and densities. Schmelz, Saba, and Strong
(1992) suggested that the Fe XVII (upper-level $2p^53d ^1P_1$) line at 15.01 ~\AA~could be used
as a measure for resonant scattering (and thus density) in the emitting plasma. Specifically the 
ratio to its Fe XVII resonance
line neighbor at 15.26~\AA~($2p^6 ~~ ^1S_0 - 2p^5 3d ^3D_1$) is of interest
(see also Waljeski et al. 1994, Brickhouse et al. 2000 and a discussion in Huenemoerder, Canizares, and Schulz 2001).
From the APED database we deduce a theoretical ratio of 3.57 for the optically thin limit. The measured ratio
in the spectrum is 3.47$\pm$1.21 and thus despite the large error bar indicates agreement with a
low density plasma. The formation radius of this line can be estimated from its line width. 
This line was indeed resolved with a HWHM of 460 km s$^{-1}$. The standard law of velocity 
for the acceleration 
zone of the wind $v(r)=v_{\infty}(1-r_{star}/r)^{\beta}$ (Lamers $\&$ Cassinelli 1999) with the index $\beta$ =0.88
and a terminal velocity of 1000 km s$^{-1}$ would then locate the line emitting region to slightly more 
than half a stellar radius from the photosphere. 

Another line that was resolved was the O VIII line at 18.97 ~\AA. Its HWHM corresponds to
velocity of 850 km s$^{-1}$, which would place the emitting region almost near the 
terminal velocity of the wind at about 7 seven stellar radii from the photosphere. 
In contrast,  the line width limit of 110 km s$^{-1}$ would place the emitting radius of 
all the other (non He-like) lines to within 10$\%$ of a stellar radius above the photosphere
(see also Cohen et al. 2002 for $\tau$ Sco). Whether there should be
an asymmetry of the line due to occultation seems irrelevant since we do not resolve most of the lines.

The forbidden lines in the He-like triplets are meta-stable and their ratio with
the corresponding intercombination lines is density sensitive above some ion-dependent theshold.
Figure~9 shows the observed He-like triplets from S XV, Si XIII, and Mg XI.
The model fits for these triplets have been optimized for each element to specifically fit the resonance line.
One of the most striking effects seen in the
triplets from O VII (very faint), Ne IX, and Mg XI (Figure~9, bottom) is that the forbidden line is not observed,
while it still appears quite prominent in Si XIII and S XV (Figure~9 middle and top).
Although the forbidden transition in the Si XIII triplet is blended with the Lyman $\gamma$ line of Mg XII,
based on the flux observed in the Mg XII Lyman $\alpha$ line, the Lyman $\gamma$ contribution cannot be
more than 5$\%$.
However we observe a forbidden line flux comparable to the expectation of the
model. We observe a similar picture in S XV (which is not blended),
however the f/i ratio here is not sensitive to densities lower than $10^{14}$ cm$^{-3}$.

All the triplets are subject to UV excitation and it has been shown by
Kahn et al. (2001) (see also Blumenthal, Drake, and Tucker 1972)
in the example of $\zeta$ Pup that the low $f/i$ ratios in line driven wind
plasmas are not likely an effect of collisional excitation in
high density plasmas, but rather are due to the destruction of the forbidden line by
the large UV flux at the corresponding excitation wavelengths. Waldron and Cassinelli (2001),
Cassinelli et al. (2001) and Miller et al. (2002) use model UV fluxes (e.g. Chavez, Stalio, and
Holberg 1995) to estimate radial constraints on the X-ray emitting plasma. Such
an analysis in the case of $\Theta^1$ Ori C is highly delicate.
One critical item is the modeling of the actual UV flux between $\sim$ 900 and 1500 ~\AA, which provides
the source for the photoexcitation in Si XIII, and Mg XI. We checked
the ratio of the UV spectra of $\Theta^1$ Ori C and  $\zeta$ Pup using Copernicus data (Snow $\&$ Jenkins 1977)
and after correcting for the different stellar parameters (19 \Rsun, 42500 K, $E_{B-V}$ = 0.07 for
$\zeta$ Pup; 8 \Rsun, 39000 K, $E_{B-V}$ = 0.35 for $\Theta^1$ Ori C; Pauldrach et al. 1994, Howard $\&$ Prinja 1989,
Bergh\"ofer, Schmitt, and Cassinelli 1996) we could not quite reconcile the model with the data in a sense
that the UV flux in $\Theta^1$ Ori C seems lower than the model.
A proper correction for extinction is certainly critical.

Effectively there no obvious reason why the UV field should be weaker than one would expect from
blackbody model atmospheres (MacFarlane et al. 1993, Chavez, Stalio, and Holmberg 1995). In this respect  
we estimated the formation radius by assuming a blackbody spectrum 
of 39000 K for the radiation field for an O7V star (Howard $\&$ Prinja 1989)
as an upper limit. The surface temperature of $\Theta^1$ Ori C
is also a source of uncertainty, as $\Theta^1$ Ori C is classified somewhere between O6 and O7.5, sometimes
even as O4. In case of the latter the result can differ by over 60$\%$. 
We calculated the photoionization (PE) rates using the recipe provided by Kahn et al. (2001, see also
Mewe $\&$ Schrijver 1978)
and obtained 2 $^3S - 1 ^1S$ decay rates from Drake (1971). In Mg XI the forbidden line can be only marginally
detected above the 1 $\sigma$ error of the continuum and here we consider the radius where the PE rate
is of the order of the decay rate as critical. The $f/i$ ratios in Si and S are, though reduced from
the one expected from the atomic data in the optically thin limit, significantly larger. Here we
have to assume that the formation radius is farther away from the stars surface and the PE rate 
is correspondingly
smaller. Thus for Mg we find an upper limit to the formation radius of 4.2 stellar radii, 
for Si 2.0 stellar radii
and 1.3 stellar radii for S.  

On the other hand it should be mentioned that in the case of zero UV flux the $f/i$ ratio of the triplets
can also be used as density diagnostics. We then obtain densities
of $<4\pm2 \times 10^{13}$ cm$^{-3}$ for Mg and $<9\pm8 \times 10^{13}$ cm$^{-3}$ for Si, repeating
the result stated in paper~I.

\begin{figure*}
\includegraphics[width=15cm]{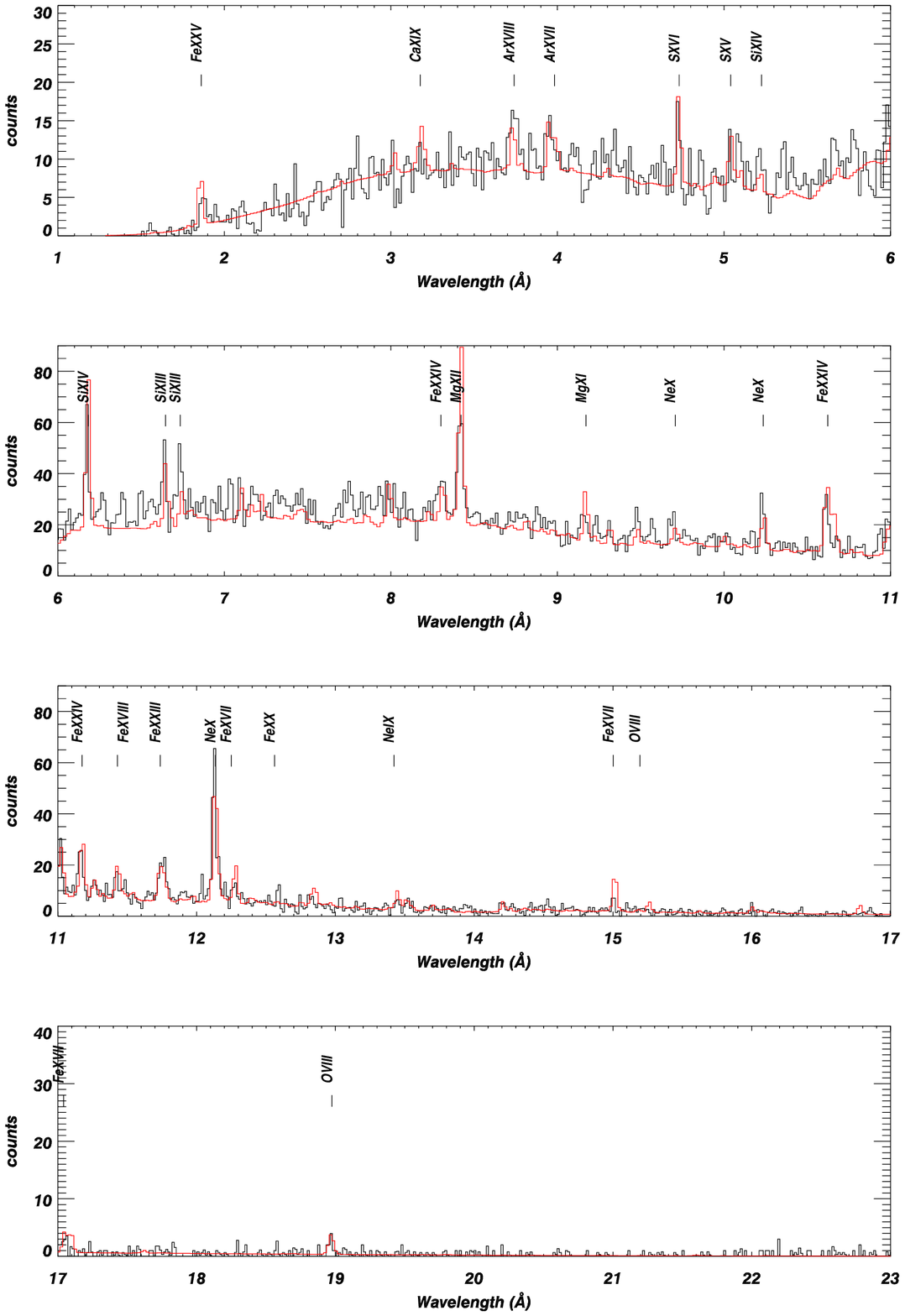}
\figcaption{The measured count spectrum for $\Theta^1$ Ori E. The red line is the best fit
found using a multi temperature APED model  }
\label{figure10}
\end{figure*}

\begin{figure*}
\includegraphics[width=15cm]{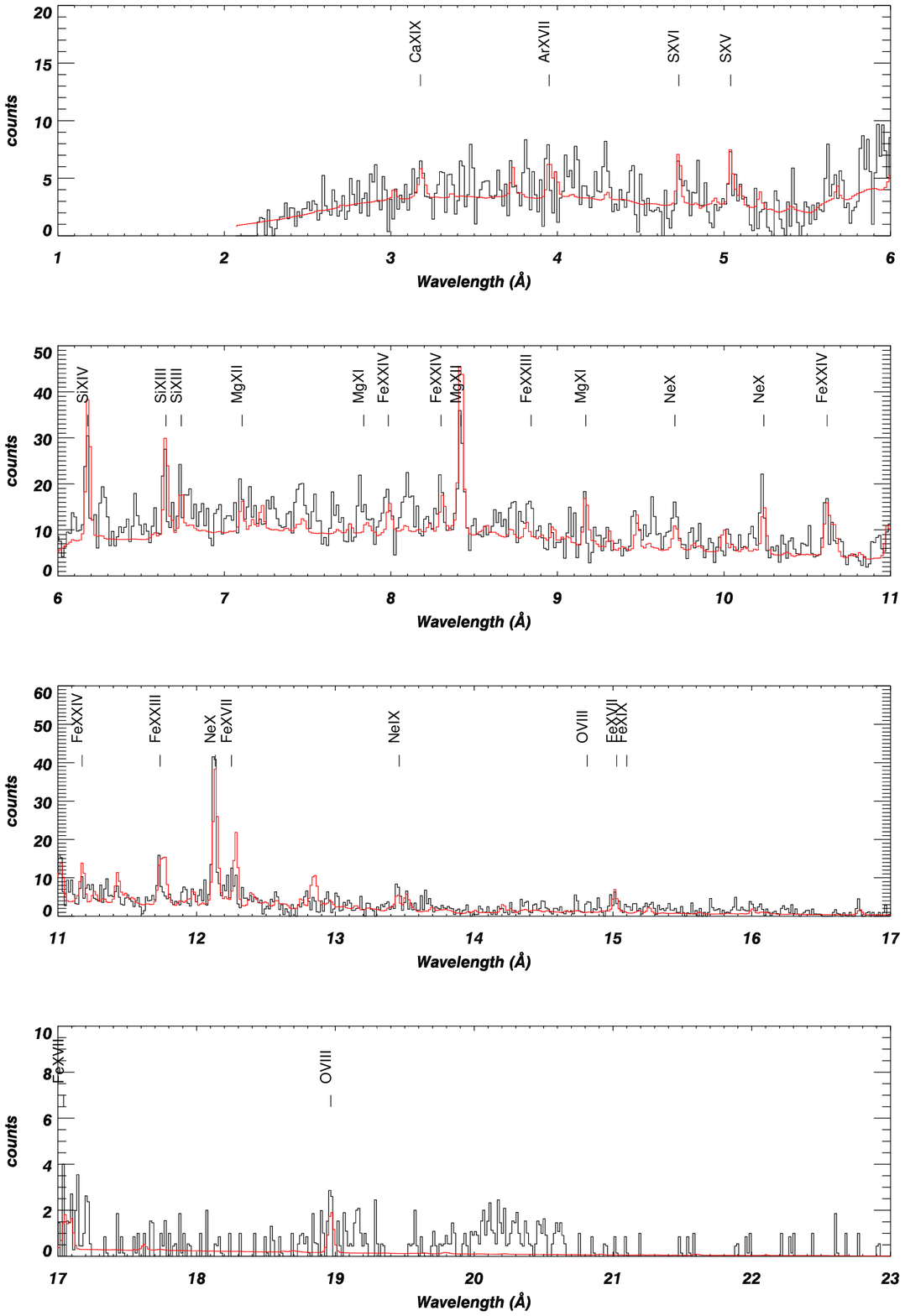}
\figcaption{The measured count spectrum for $\Theta^1$ Ori A. The red line is the best fit
found using a multi temperature APED model  }
\label{figure11}
\end{figure*}

\subsection{$\Theta^1$ Ori E and A}

The next two X-ray brightest stars in the Trapezium are stars A and E with X-ray fluxes of 1.3 to 
2.5$\times 10^{-12}$ \ergcm
corresponding to luminosities of 2 to 4$\times 10^{31}$ \ergsec, respectively (Schulz et al. 2001). 
These fluxes are an order of magnitude fainter than $\Theta^1$ Ori C, which is likely due to 
their B spectral types (Cassinelli et al. 1994).
In this respect the HETGS spectra are less brilliant, however in both
spectra we are still able to detect quite a number of emission lines and strong continua. 
This at least allows us to sufficiently constrain our multicomponent model. Table~3 shows the
characteristics of the brightest lines for both stars. These line properties are very similar to 
the characteristics exhibited by $\Theta^1$ Ori C. 
This is specifically noteworthy for the Fe XXIV transitions.
The lines appear unresolved in both stars and we can set a HWHM confidence limit of 160 km s$^{-1}$.
In general we observe the same  
most remarkable features: the strong continuum and very narrow and symmetric lines.

Figures 10 and 11 show the count spectra of the two stars binned to 0.015~\AA. The red solid
lines are the final model spectra.
We fitted these spectra following the procedure described in section 3.2.2. The result
for $\Theta^1$ Ori E is shown in the middle part, for $\Theta^1$ Ori A in the bottom
part of Table 3. Both spectra accepted 5 components including a quite marginal low temperature
component. The four higher temperature components seem similar to components 2-5 in
$\Theta^1$ Ori C, with the one for the very high temperature missing.
In this respect we observe a similar bifurcation in emissivity
in stars A and E, which includes low temperature emissivity below at or below 10$^7$ K
and strong emissivity that has temperatures significantly higher than 10$^7$ K.
In the case of $\Theta^1$ Ori E roughly 80$\%$ and
for $\Theta^1$ Ori A 70$\%$ of the luminosity appears at temperatures higher than 1.5$\times10^7$ K.
Over 50$\%$ of the luminosity originates from emitting plasmas showing temperatures
higher than 3$\times10^7$ K. Thus if we again only count the flux of the low temperature
component as the one form the unconfined wind, the ratio $L_x^{wind}/L_{bol}$ for these two sources
are similar the the one for $\Theta^1$ Ori C and thus consistent with normal O-stars.
The fit abundances are also similar to values for $\Theta^1$ Ori C,  except
that the Ne lines are stronger, and we do not detect a deviation from solar values.

Most He-like triplet are too weak to perform a meaningful analysis except for Si XIII,
where in the case of star E (Figure 10, 6.74~\AA) we observe a forbidden line that
is almost as strong as the resonance line. In star A it is not as strong but clearly
detectable.
In both cases we do not detect an intercombination line and by setting
a limit to the line using the 1 $\sigma$ error of the continuum we can only get rough
estimate of the $f/i$ ratio.
We nevertheless we can put some limits on the formation radius of Si XIII.
In the optically thin limit without photoexcitation we expect 2.16 for the $f/i$ ratio.
For star E we can limit the ratio to 3.43$\pm$1.69, consistent with the then limit
and basically no UV destruction. Assuming a B0.5V star with a surface temperature
of 23000, we would expect the forbidden line would get completely destroyed
within 0.4 stellar radius above the photosphere. Since this is not the case
and the line seems to be fully intact, we assume that the  2 $^3S - 1 ^1S$ decay rate
entirely dominates the PE rate, which should occur at a distance of 1.4 stellar radii.
For star A the ratio is 2.13$\pm$1.33 we can make the same argument assuming a surface
temperature of 20000 K (B1V, Strickland 2000). In terms of density, i.e. under the assumption
that there is no photoexcitation, the $f/i$ ratios seem to be in compliance with the
densities below 10$^{13}$ g cm$^{-3}$.

\vspace{0.3cm}
\vbox{
{\sc TABLE~3 : IDENTIFIED IONS IN $\Theta^1$ Ori A and E}
\vskip 4pt
\begin{tabular}{lccc}
\hline
\hline
 ion & $\lambda_o$ & flux ($\Theta^1$ Ori A)  & flux ($\Theta^1$ Ori E) \\
     & \AA  & 10$^{-6}$ ph s$^{-1}$~cm$^{-2}$ & 10$^{-6}$ ph s$^{-1}$~cm$^{-2}$ \\
\hline
 & & & \\
Fe XXV &  1.861 &    -  &   7.182$\pm$  1.077 \\
Ca XIX &  3.198 &   0.895$\pm$  0.134 &   1.481$\pm$  0.222 \\
Ar XVIII &  3.734 &   -  &   3.321$\pm$  0.498 \\
Ar XVII &  3.949 &   0.298$\pm$  0.045 &   2.274$\pm$  0.341 \\
S XVI &  4.730 &   0.840$\pm$  0.126 &   5.233$\pm$  0.785 \\
S XV &  5.039 &   1.097$\pm$  0.165 &   2.737$\pm$  0.411 \\
Si XIV &  5.217 &   -  &   1.194$\pm$  0.179 \\
Si XIV &  6.183 &   3.293$\pm$  0.494 &   3.935$\pm$  0.590 \\
Si XIII &  6.648 &   1.731$\pm$  0.260 &   2.573$\pm$  0.386 \\
Si XIII &  6.740 &   1.706$\pm$  0.256 &   2.751$\pm$  0.413 \\
Fe XXIV &  7.989 &  0.676$\pm$  0.101 & - \\
Fe XXIV &  8.304 &    0.676$\pm$  0.101  &   1.658$\pm$  0.249 \\
Mg XII &  8.420 &   2.612$\pm$  0.392 &   5.060$\pm$  0.759 \\
Fe XXIII &  8.815 & 1.363$\pm$  0.204 & - \\
Mg XI &  9.169 &   1.963$\pm$  0.294 &   2.495$\pm$  0.374 \\
Ne X &  9.708 &   0.811$\pm$  0.122 &   2.393$\pm$  0.359 \\
Ne X & 10.239 &  3.078$\pm$  0.462 &   3.530$\pm$  0.530 \\
Fe XXIV & 10.619 &  2.919$\pm$  0.438 &   3.508$\pm$  0.526 \\
Fe XXIII & 10.983 &  - &   2.048$\pm$  0.307 \\
Fe XXIV & 11.176 &  1.013$\pm$  0.152 &   4.392$\pm$  0.659 \\
Fe XXIII & 11.736 &  2.186$\pm$  0.328 &   2.875$\pm$  0.431 \\
Fe XVIII & 11.330 &  - &   3.442$\pm$  0.516 \\
Ne X & 12.135 &  14.589$\pm$  2.188 &  16.331$\pm$  2.450 \\
Fe XVII & 12.261 &  1.558$\pm$  0.234 &   0.498$\pm$  0.075 \\
Fe XX & 12.576 &  -  &   1.617$\pm$  0.242 \\
Ne IX & 13.445 &   5.727$\pm$  0.859 &  37.722$\pm$  5.658 \\
Fe XVII & 15.014 &  21.573$\pm$  3.236 &   4.164$\pm$  0.625 \\
Fe XIX & 15.079 &  8.305$\pm$  1.246 & - \\
O VIII & 15.176 &  - &  18.360$\pm$  2.754 \\
Fe XVII & 17.051 &  5.309$\pm$  0.796 &  13.059$\pm$  1.959 \\
O VIII & 18.970 &  0.177$\pm$  0.027 &   9.362$\pm$  1.404 \\
 & & & \\
\hline
\end{tabular}
\normalsize
}
\vspace{0.3cm}

\section{Discussion}

High resolution X-ray spectra provide new and more powerful diagnostics of
the high energy emission from hot star plasmas. These diagnostics involve line identifications,
the relation of specific line ratios with physical parameters such as temperature and
density, line shapes and shifts as well as full scale plasma models. 
In paper~I we presented a preliminary line list for $\Theta^1$ Ori C. 
For this paper we added one more observation to the analysis for a net exposure of 83 ks.
The availability of improved calibration products and models allowed us to refine 
these results in major areas.
For $\Theta^1$ Ori C the number of detected lines increased and allowed for better  
constraints on the plasma modeling specifically for the calculation of the DEM.  
The DEM analysis of the phase averaged spectrum as well as the fitting
of the broad band plasma model showed that the bulk of the X-ray emission comes from 
plasmas with temperatures above 15 Million K
and only a small, well separated part of the emission corresponds to emission of lower than
10 Million K. 

There are models related to the standard line-driven wind instability model
that are able to allow temperatures in excess of 10 Million K (Feldmeier et al. 1997a,
Howk et al. 2000, Runacres$\&$Owocki 2002). However possible high shock velocities
near the onset of the wind cannot produce enough volume emissivity to account for the 
X-ray spectrum. The results also confirm a similar behavior for $\Theta^1$ Ori A $\&$ E. 
This means the bulk of the X-ray emission in the bright Orion Trapezium stars
is not compatible with any form of instabilities in a line-driven wind. In fact these stars
seem to be of a hybrid nature where only a small fraction of the X-rays are produced in wind 
shocks, a larger fraction shows a magnetic origin and bears striking similarities 
to the hard X-ray emission pattern observed in stars with active coronae (Huenemoerder,
Canizares, and Schulz 2001, Huenemoerder et al. 2003). 

The DEM model allows for a non-thermal continuum component that contributes less than
1$\%$ to the total flux for a power law index of -0.5 as suggested by Chen and White (1991).
These authors proposed a non-thermal origin for possible hard X-rays in form of
an inverse Compton continuum. This limit corresponds to 4$\times10^{30}$ \ergsec, which is an
order of magnitude higher than predicted. Thus we may not really be sensitive to the issue.
It is quite clear though that the high energy flux from the Orion Trapezium stars
is not due to a possible non-thermal origin.

\subsection{Possible Shocked Wind Component}

The emissivity peak at
log T = 6.9 includes most lines from O to Si and ions below Fe XX.    
These are the moderate temperature lines we also observe in wind shocks 
in $\zeta$ Pup and others 
(Cassinelli et al. 2001, Kahn et al. 2001, Waldron and Cassinelli 2001, Schulz et al. 2001a,
Schulz et al. 2002) suggesting a similar origin for this component on $\Theta^1$ Ori C. 
This peak does not change within the two observed phases of the 15.4 day
cycle in $\Theta^1$ Ori C.  
Ignace (2001) and Owocki and Cohen (2001) calculated the shape of X-ray
line profiles in stellar winds under various conditions.  

The fact that most lines in the spectra do not show
significant shifts and are 
unresolved puts limits on these conditions. If the lines were to be produced at large
radii in the outer wind, we should observe broad, symmetric, sometimes flat topped lines.
Yet for wind shocks, the only way to achieve symmetry is to have almost no attenuation in wind.
Given the low mass loss rate of $\sim 4\times 10^{-7}$ \Msun yr$^{-1}$ (Howarth $\&$ Prinja
1989) this is quite likely.
Some of the  X-ray emitting plasma also has to be near the photosphere at the onset of the wind
given the unresolved nature of the lines. However it seems that some of the softest lines
are indeed resolved and show velocity broadening of up to 850 km s$^{-1}$. This is indicative  
that these lines contributing to the low temperature emissivity
peak are due to X-rays from shocks in the outer wind. Would 
this low temperature emissivity be the only 
contribution to the X-ray emission, the Orion stars would be quite X-ray faint, 
but with a log $L_x/L_{bol}$ between -7.2 and -7.6, which is near the canonical value for 
O stars. 
  
\subsection{Magnetically Confined Winds}

The truth is, however, that log $L_x/L_{bol}$ is more of the order of -6.5 for these stars.
Since most emissivity appears to originate at temperatures that are 
incompatible with wind shocks, we have to look for other mechanisms. 
There are several indications that the enhanced X-ray activity could be triggered by 
magnetic fields. Gagne et al. (1997) interpreted the strong 15.4 day period in the 
X-ray and optical emission of $\Theta^1$ Ori C reported by Stahl et al. (1996)
in terms of the star's significant magnetic field.

Babel and Montmerle (1997a) proposed a magnetically confined wind shock (MCWS) model based on an
oblique magnetic rotator model for $\Theta^1$ Ori C. Here the wind is confined by
a magnetic dipole field and forced into the magnetic equatorial plane. The observed
15.4 day variability thus is produced by the tilt of the dipole field relative to the 
rotational axis. The predicted field is of the order of 300 G. 
Very recent spectrophotometric observations indicate a dipole field of 1.1 kG with
an inclination of 42$^{\rm o}$ with respect to the stars rotational axis (Donati et al. 2002).
These values are 
quite consistent with the temperatures we observe in the spectrum 
of $\Theta^1$ Ori C. This model
successfully explains the periodicity in the H$\alpha$ and H II lines as well 
as in P Cygni line profiles in the UV (Stahl et al. 1993, Walborn $\&$ Nichols 1994, 
Stahl et al. 1996, Reiners et al. 2000). 
In the X-ray light curve this interpretation of the periodicity is also attractive.

The HETG spectra, however, indicate a more complex behavior. The HETG observations
were performed at phases around 0.37 and 0.82.
According to the \ros HRI lightcurve in Gagne et al. (1997) we should see a difference
in flux between the two observations of about 30$\%$. This applies to the
non-varying low temperature emissivity peak and we see the rise towards the first 
high temperature peak, which includes all the lines that were accessible to 
the \ros bandpass. The HETG data show that there is more going on than just a flux change.
We observe a dramatic change of emissivity 
at log T (K) = 7.4 between the two phases. Donati et al. (2002)
point out that meaningful comparisons with predictions of the MCWS model can
only be achieved from lines at extreme configurations (phases 0.0 and 0.5).   
Our phase should be somewhere in between these two extremes.
The model  
states that the change in luminosity between the two phases above 2 keV (below $\sim$ 6\AA) 
should decrease compared to below 2 keV (above $\sim$ 6\AA). 
This is not what we observe. The difference below 6\AA~ is largest
instead with almost 40$\%$. Furthermore we see no obvious reason in the MCWS, 
why the plasma temperature distribution
should change like we observe in the DEM.   

More quantitative modeling including a more detailed energy balance treatment 
by Ud'Doula and Owocki (2002) allow us to further differentiate the phenomenology
between a magnetic field and a stellar wind. By relating the magnetic energy
density to the matter outflow in the wind a confinement parameter 
$\eta \sim {B R^2}/{\dot M v_{\infty}}$ can be defined relating the magnetic field density
at a radius R from the star and the mass loss rate $\dot {\rm M}$ in a stellar
wind of terminal velocity v$_{\infty}$. Their magnetohydrodynamical simulations
showed that in the case of a large $\eta$ value ($\sim 10$) and closed magnetic field
topologies near the star the wind is forced into loop-like
structures in which strong shock collisions produce hard X-rays.
Ud'Doula and Owocki (2002) showed that these structures could generate enough emissivity at high
temperatures by applying the standard shock jump condition from Babel and Montmerle (1997b). 
For observed values of magnetic field strength,
terminal wind velocity, and mass loss rates for $\Theta^1$ Ori C, the $\eta$ parameter
is well over 10 (Gagne et al. 2001). Donati et al. (2002) applied the standard model devised by 
Babel and Montmerle (1997b) and found similar trends in terms of average temperature and density. 

The loop-like structures proposed by Ud'Doula and Owocki (2002) and
Gagne et al. (2001) may also 
explain the structure we observe in the DEM distribution where different
peaks may relate to different confined structures. 
We can further
speculate that changes at different phases refer to different
structures. These confined structures may 
be quite unstable on short time scales resulting in highly variable 
X-ray emission.  
Feigelson et al. (2002) recently detect rapid variability in the 
O7pe star $\Theta^2$ Ori A and suggested unseen companions for these
stars (see also below) . This may not be necessary. As an interesting analogy, the DEMs recently 
deduced from X-ray emission form active coronae in cool stars like AR Lac (Huenemoerder
et al. 2003) and II Peg (Huenemoerder, Canizares, and Schulz 2001) appear very
similar during flares, where in addition to a small low temperature peak one or more
strong high temperature peaks evolve. The nature of the spectra in these cases 
are of striking similarity to the ones we observe in the Orion Trapezium
with strong continua and unresolved lines. In this respect we may speculate
that in the case of the Orion stars matter is constantly supplied into magnetic 
loops by the wind generating  shock jumps on a permanent basis. In other 
words, these stars would be in a permanent state of flaring. The 
means of energy deposit is expected to be different in cool stars, which
posses a dynamo and deposit energy into the plasma via reconnection.
The mechanism in the winds of hot stars is more indirect via confining.
Clearly, these details still have be worked out.  

If these shocks are magnetically confined they are not likely to produce 
line shifts and significant line broadening. Donati et al. (2002) simulated 
dynamic X-ray line spectra for several model cases for $\Theta^1$ Ori C  based
on the MCWS model and found narrow line shapes at the observed the phases.
Any predicted line shifts are of the order of 150 to 200 km s$^{-1}$ and below our 
resolving power. 
This is consistent with the line characteristics 
in all three Orion spectra. 
 
\subsection{Low-Mass Companions}

The Trapezium stars are known to have one or more companions. Quite recently Weigelt et al. (1999)
reported on the existence of a close, probably low mass companion to 
$\Theta^1$ Ori C and confirmed the companion
detected by Petr et al. (1998) in $\Theta^1$ Ori A. The companion in $\Theta^1$ Ori C is as close
as 33 milli-arcseconds. The close companion in $\Theta^1$ Ori A has a separation of
202 milli-arcseconds.   
The only star in the Trapezium with no detected companion is $\Theta^1$ Ori E.
Based on the median age of the cluster of 0.3 Myr (Hillenbrandt 1997) and the ubiquity of nearby
proplyds (O'Dell, Wen, $\&$ Hu 1993, Bally et al. 1998), which likely contain Class II T Tauri stars
(Felli et al. 1993, McCaughrean $\&$ Stauffer 1994, Schulz et al. 2001), we consider
these companions to be young T Tauri stars.  Weigelt et al. (1999) similarly suggest
that the companion in $\Theta^1$ Ori C is a very young intermediate- or low-mass (M $<$ \Msun)
star based on evolutionary pre-main sequence (PMS) evolutionary tracks. 

The X-ray emission of most young PMS stars in the Orion Trapezium Cluster is
usually absorbed (Garmire et al. 2000), but emit hard 
X-ray emission with temperatures of around
30 Million K (Schulz et al. 2001). Giant X-ray flares can reach up to 60 to 100 Million K 
(Kamata et al. 1997, Yamauchi $\&$ Kamimura 1999, Tsuboi et al. 2000).
X-ray contribution from such flares can be ruled out for $\Theta^1$ Ori C, A, and E by the fact that we 
do not observe variability in the light curve (here we do not take
in account high-frequency variability as observed by Feigelson et al. 2002).
If we see contributions from the low-mass
companion, it has to be persistent. X-ray luminosity functions from many star forming regions
peak at luminosities below 10$^{31}$ \ergsec. This has also been been observed for the  
low-mass PMS population in the vicinity of the Orion Trapezium (Schulz et al. 2001, Feigelson et al. 2002).
In this respect a major contribution from such a companion to the spectrum 
in $\Theta^1$ Ori C can be ruled out. 

It is possible that X-rays from a low-mass companion make a significant 
contribution to $\Theta^1$ Ori A, since its
flux components are an order of magnitude fainter. Even if there were such 
a contribution, it can only be a small fraction of the total observed X-ray luminosity. This is even more 
the case for $\Theta^1$ Ori E should
it harbor an unseen companion. We can thus rule out that the X-ray emissivity pattern
is due to a possible binary nature of the stars.

\subsection{Evolutionary Implications}

The very similar properties and morphologies in the spectra of $\Theta^1$ Ori A, C, and E 
raises an intriguing issue.
With an ionization age of the nebula of about 0.2 Myr and a median age of the cluster 
of about 0.3 Myr it is quite suggestive that the Trapezium stars 
are true ZAMS stars. Zero-age here is considered to be the time when energy generation by nuclear
reactions first fully compensates the energy loss due to radiation from the stellar photosphere. 
In the case of $\Theta^1$ Ori C it has been stated many times, that the magnetic activity 
could possibly be of pre-main sequence origin (Gagne et al. 2001, Donati et al. 2002). 
Gagne et al. (2002) noted that 
stars like $\Theta^1$ Ori C, or $\tau$ Sco all show signs of magnetic activity and that they
are all associated with young star-forming regions. With the Orion Trapezium we may have 
an indication that magnetic fields may be anti-correlated with age. Four out of the five main 
Trapezium stars now show
strong magnetic signatures, with $\Theta^1$ Ori A, C, and E the most striking cases. So far the 
view of primordial high fields has been limited to peculiar (Ap and Bp) stars and its has long been 
suspected that 
$\Theta^1$ Ori C is a candidate for an Op star (Gagne et al. 2001).   
$\Theta^1$ Ori D, so far also classified as a peculiar dwarf (B 0.5Vp), is weak in X-rays. 
The X-ray spectrum (Schulz et al. 2001)
does not indicate very high temperatures and the grating data are too marginal to search for narrow lines. 
However,  it seems that for some reason all Trapezium stars are chemically peculiar either by coincidence or this
peculiarity has something to do with magnetism and/or their young age.

\vspace{0.3cm}
\begin{table*}
{\sc TABLE~4 YOUNG MASSIVE STARS VS. MAGNETIC ACTIVITY}
\vskip 4pt
\begin{tabular}{lccccccccc}
\tableline
\tableline
star&spectral&age&Star Form.&T &Magn. &log L$_x^{mag}$ &log L$_x^{wind}$ &log L$_x^{wind}/L_{bol}$ &refs.\tablenotemark{a} \\
    & type   &My & Region   &MK&Fields&\ergsec         & \ergsec          &                          &  \\
      &          &       &         &     &  & & & &         \\
\tableline
      &          &       &         &     &  & & & &        \\
$\Theta^1$ Ori A & B0.5V & 0.3 & Orion  & 5-43 & yes & 31.0 & 30.7 & -7.3 & 1,2,3\\
$\Theta^1$ Ori B & B1V/B3 & 0.3 & Orion & 22-35 & prob. & 30.3 & ? & ? & 2,3,4\\
$\Theta^1$ Ori C & O6.5Vp & 0.3 & Orion  &6-66 & yes & 32.2 & 31.5 & -7.2 & 1,2,3\\
$\Theta^1$ Ori D & B0.5Vp& 0.3 & Orion  &7-8  &  ? & ? & 29.5 & -8.5 &2,3,4\\
$\Theta^1$ Ori E & B0.5  & 0.3 & Orion  & 4-47 & yes & 31.4 & 30.8 & -7.2 &1,2,3\\
$\Theta^2$ Ori   & O9.5Vpe& 0.3 & Orion & 5-32 & yes & 31.1 & 31.4 & -7.1 &7,12\\
$\tau $ Sco      & B0.2V & $~\sim$1 & Sco-Cen & 7-27 & yes & 31.9 & 31.4 & -7 & 5,6\\
HD 164492A       & O7.5V & 0.3 & Trifid  & 2-12? & ? & ? & 31.4 & 7.1 & 8,8a \\
HD 206267        & O6.5V & 3-7 & IC 1396 & 2-10 & no & -  & 31.6 & -7.7 & 9,10\\
$\tau$ CMA       & O9 Ib & 3-5 & NGC 2362 & 3-12 & no & - & 32.3 & -7.2 & 9\\
15 Mon           & O7V   & 3-7 & NGC 2264 & 2-10 & no & - & 31.7 & -7.2 & \\
$\iota$ Ori      & O9 III& $<$12  & Orion &  1-10 & no & -  & 32.4 & -6.8 & 9\\
$\zeta$ Ori      & O9.7 Ib & $<$12  & Orion &   & no & - & 32.5 & -6.8 & 11\\
$\delta$ Ori     & O9.5II  & $<$12  & Orion &   & no & - & 32.2 & -6.8 & 13\\
      &          &       &         &     &  & & & &         \\
\tableline
\tablenotetext{a}{
(1) this paper; (2) Hillenbrandt (1997); (3) Petr et al. (1998); (4) Schulz et al.
(2001); (5) Cohen, Cassinelli $\&$ Waldron (1997);  (6) Killian (1994);
(7) Schulz et al. (2003a); (8) Rho et al. (2001); (8a) J. Rho, priv. comm.; (9) Wojdowski et al. (2002);
(10) Strickland et al. (1997); (11) Waldron $\&$ Cassinelli (2001); (12) Feigelson et al. (2002);
(13) Miller et al. (2002) }
\end{tabular}
\end{table*}

\vspace{0.3cm}

In Table 4 we
list some properties from presumably young massive stars. These properties include
the spectral type, the assumed cluster age, star forming region, and the 
currently known X-ray temperatures. The compilation is certainly far from complete,
but demonstrates that stars from regions younger than 1 Million yr 
show clear evidence for magnetic activity. The main  
identifier for magnetic activity is the extremely high ($> 10^7$ K) persistent temperature.
For the Orion stars and $\tau$ Sco we additionally observe symmetric and narrow lines.
HD 164492 is an interesting case as it is at the center of a very active
star-forming region associated with the Trifid nebula, which very recently has been
classified to be in a "pre-Orion" evolutionary stage (Lefloch $\&$ Cernicharo 2000) and
thus maybe the "youngest" massive ZAMS star in the sample. Although not fully resolved, its
X-ray characteristics (Rho et al. 2001) as observed with \asca seem similar to
what has been observed for $\Theta^1$ Ori C with \asca. Stars like HD 206267 and $\tau$ CMa are
at the center of more evolved clusters and do not show these characteristics. In fact,
they behave very much like $\zeta$ Pup (Kahn et al. 2001, Cassinelli et al. 2001) in that
they show temperatures and line shapes consistent with shock instabilities in a 
radiation driven wind (Wojdowski et al. 2002, Schulz et al. 2001a, Schulz et al. 2002).  
Finally $\iota$ Ori and $\zeta$ Ori are part of the Orion region which is is older than
the Trapezium cluster core and further evolved. Here we also see no indication for
magnetic activity (Schulz et al. 2002). $\zeta$ Ori (Waldron and Cassinelli 2001) maybe
a special case in that it lacks high temperatures but possesses symmetric X-ray lines,
which likely are due to a low opacity of the wind.

Table 4 may reflect an evolutionary sequence in which massive stars that
enter the main sequence carry strong magnetic fields interacting with an emerging wind.
In the table we list two types of X-ray luminosities based on the dichotomy observed
in the Orion stars, L$_x^{mag}$ for the amount produced by presumably magnetic confinement and
L$_x^{wind}$ for the amount from wind shocks. The bolometric luminosities for the 
$L_x^{wind}/L_{bol}$ ratios were taken from Bergh\"ofer, Schmitt, and Cassinelli (1996) 
Once the star evolves, magnetic fields may become less important as mass loss rates are higher.  
The X-ray emission sooner or later are fully generated by instabilities in the wind only.
Here studies that search for
magnetic fields in hot stellar winds (see Chesneau $\&$ Moffat 2002) are useful.
An intriguing aspect on the Orion stars is that there also seems to exist relatively
faint X-ray emission consistent with non-magnetic wind shocks. Thus even if it is the case
that most Orion stars are indeed just abnormal and strong magnetic fields are more
the exception than the rule, then very young massive stars if not magnetic should at least scale 
with their bolometric luminosity with $L_x^{wind}/L_{bol} \sim -7$ and not higher.
This may explain the relative X-ray weakness
of HD 164492 A relative to $\Theta^1$ Ori C (J. Rho, private communication), although
$\Theta^1$ Ori C may be even younger than HD 164492 A.
This may not explain the extremely weak X-rays flux from $\Theta^1$ Ori D, though. From
a projected bolometric luminosity for a B0.5V star of 10$^{38}$ \ergsec and an
X-ray luminosity of 3$\times10^{29}$ \ergsec (from Schulz et al. 2001)
we compute a $L_x^{wind}/L_{bol}$ of -8.5 for the star, which is very low.
Om the other hand we can turn the argument around and in the case of $\tau$ Sco use
$L_x^{wind}/L_{bol}$ = -7 to estimate the of amount of L$_x^{mag}$ out of the observed ratio as
has been in done in Table 3. 

Too little is known about massive stars 
as they quickly evolve towards the main sequence. Although it is well accepted that
many high-energy processes in YSOs are dominated by magnetic activity (see
Feigelson $\&$ Montmerle 1999 for a review) and hard X-rays in flares are produced by
powerful magnetic reconnections, star-disc activities and jet-formation, these
findings are still confined to low mass stars. 
In several models for class I and II YSOs magnetic field play a crucial role
in the X-ray production, either in form of X-ray winds and accretion (Shu et al. 1997)
or accretion shocks (Hartmann 1998).
It has yet to be established how these models for low-mass YSOs may relate
to more massive stars. 

One should be careful, however, to simply argue with age in such a correlation.
It has to be kept in mind that there are similar patterns once
we include mass loss rate and wind velocities. In this respect the stars associated with
magnetic activity also have low mass loss rates, low wind velocities and many of them
are optically classified as dwarfs. All
these properties have to be accounted for.  Future studies
need to incorporate a large number of highly resolved X-ray spectra of these young
massive stars in correlation with fundamental stellar wind properties to 
establish an evolutionary pattern that leads from either faint or magnetically dominated
X-ray sources to strong wind instability dominated X-ray sources.   

\acknowledgements

We thank all the members of the \chandra team for their enormous efforts and the referee for
valid comments to improve the menuscript. This research is funded
in part by the Smithonian Astrophysical Observatory contract SV-61010 (CXC) and NAS8-39073 (HETG) 
under the Marshall Space Flight Center.


\begin{references}
\reference{bab1997} Babel J., $\&$ Montmerle T., 1997b, \aap~323, 121
\reference{bab1997a} Babel J., $\&$ Montmerle T., 1997a, \apj~485, L29
\reference{bal1998} Bally J., Sutherland R.S., Devine D., and Johnstone D., 1998, \aj ~116, 293
\reference{ber1994} Bergh\"ofer T.W., and Schmitt J.H.M.M., 1994, \aap~292, L5
\reference{ber1996} Bergh\"ofer T.W.,  Schmitt J.H.M.M., and Cassinelli J.P, 1996, \aap ~31, 289
\reference{blu1972} Blumenthal G.R., Darke G.W.F., and Tucker W.H., 1972, \apj~172, 205
\reference{boh1981} Bohlin R.C. \& Savage B.D., 1981, \apj ~249, 109
\reference{bri2000} Birckhouse N.S., Dupree A.K., Edgar R.J., Liedahl D.A., Crake S.A., White N.E., and Singh K.P., 2000,\apj~530, 387
\reference{cas2001} Casinelli J.P., Miller N.A., Waldron W.L., MacFarlane J.J., and Cohen D.H., 2001, \apj ~554, 55
\reference{cas1994} Casinelli J.P., Cohen D.H., MacFarlane J.J., Sanders W.T., and Welsh B.Y., 1994, \apj~421, 705 
\reference{cas1983} Cassinelli J.P, and Swank J.H., 1983, \apj~ 271, 681
\reference{cha1995} Chavez M., Stalio R., and Holberg J.B., 1995, \apj~449, 280
\reference{che1991} Chen W., $\&$ White R.L., 1991, \apj~366, 512
\reference{che2002} Chesneau O. \& Moffat A.F.J., 2002, \pasp ~114, 612
\reference{chl1989} Chlebowski T., Harnden F.R. jr., and Sciortino S., 1989, \apj~ 341, 427
\reference{coh2002} Cohen D.H., de Messi\'eres G.E., MacFarlane J.J., Miller, N.A., Cassinelli J.P., Owocki, S.P., and Liedahl, D.A., 2002, \apj~, submitted
\reference{coh1997} Cohen D.H., Cassinelli J.P., and Waldron W.L., 1997, \apj~488, 397
\reference{coh1997a} Cohen D.H., Cassinelli J.P., and Macfarlane J.J., 1997, \apj ~487, 867
\reference{coh1996} Cohen D.H., Cooper R.G., Macfarlane J.J., Owocki S.P., Cassinelli J.P, and Wang P., \apj ~460, 506
\reference{cor1994} Corcoran M.F., Waldron W.L., MacFarlane J.J., et al., 1994, \apj~436, L95
\reference{don2002} Donati J.-F., Babel J., Harries T.J., Howarth I.D., Petit P., and Semel M., 2002, \mnras ~333, 55
\reference{dra1971} Drake G.W., 1971, Phys. Rev. A, 3, 908
\reference{fei2002} Feigelson E.D., Broos P., Gaffney J.A. III, Garmire G., Hillenbrand L.A., Pravdo S.H., Townsley L., and Tsuboi Y., 2002, \apj ~574, 258 
\reference{fei1999} Feigelson E.D., and Montmerle T., 1999, \araa~37, 363
\reference{fel1997} Feldmeier A., Kudritzki R.P., Palsa R., Pauldrach A.W., and Puls J., 1997, \aap~ 320, 899
\reference{fel1997a} Feldmeier A., Puls J., and Pauldrach A.W., 1997, \aap~ 322, 878
\reference{fel1993} Felli M., Taylor G.B., Catarzi M., Churchwell E., and Kurtz S., 1993, \aap Suppl. 101, 127
\reference{gag2001} Gagne, M., Cohen, D., Owocki, S., and Ud-Doula, A., 2001, in X-ray at Sharp Focus, ASP Conference Series, eds. S. Vrtilek, E.M.Schlegel, L. Kuhi
\reference{gag1995} Gagn\'e M., Caillault J.-P., and Stauffer J.R., 1995, \apj~445, 280
\reference{gag1997} Gagn\'e M., Caillault J.-P., Stauffer J.R., and Linsky J.L., 1997, \apj~ 478, L87
\reference{gar2000} Garmire G., Feigelson E.D., Broos P., Hillenbrand L., Pravdo S.H., Townsley L., and Tsuboi Y., 2000, \aj~120, 1426
\reference{hun2003} Huenemoerder D.P., Canizares C.R., Drake J.J., and Sanz-Forcada J., 2003, \apj~, submitted 
\reference{hun2001} Huenemoerder D.P., Canizares C.R., and Schulz N.S., 2001, \apj~559, 1135 
\reference{Har1979} Harnden F.R. jr., Branduardi G., Gorenstein P., Grindlay J., Rosner R., Topka K., Elvis M., Pye J.P.,
and Vaiana G.S., 1979, \apj ~234, 55
\reference{hil1993} Hillier D.J., Kudritzki R.P., Pauldrach A.W., Baade D., Casinelli J.P., Puls J., and Schmitt J.H.M.M., 1993, \aap~ 276, 117
\reference{hil1997} Hillenbrand L.A., 1997, \aj~113, 1733
\reference{hoq1989} Howarth I.D. \& Prinja R.K., 1989, \apj S ~69, 527
\reference{how2000} Howk J.C., Cassinelli J.P., Bjorkman J. E., Lamers H.J.G.L.M., 2000, \apj ~534, 348
\reference{hun2001} Huenemoerder D., Canizares C.R., and Schulz N.S., 2001, \apj ~559, 1135
\reference{ign2001} Ignace R., 2001, \apj ~549, 119
\reference{kah2001} Kahn, S.M., Leutenegger, M., Cottam, J., Rauw, G., Vreux, J.M., den Boggende, T., Mewe, R., \& Guedel, M., 2001, \aap, 365, L312
\reference{kam1997} Kamata Y., Koyama K., Tsuboi Y., and Yamauchi S., 1997, \pasj~69, 461
\reference{kil1994} Kilian J., 1994, \aap ~282, 867
\reference{Lam1999} Lamers H.J.G.L.M. and Cassinelli  J.P., 1999, Introduction to Stellar Winds, Cambridge Univ. Press, Cambridge
\reference{lef2000} Lefloch B \& Cernicharo J., \apj ~545, 340
\reference{luc1980} Lucy L.B., and White R.L., 1980, \apj~241, 300
\reference{Luc1982} Lucy L.B., 1982, \apj~ 255, 286
\reference{mac1993} MacFarlane J.J., Waldron W.L., Corcoran M.F., Wolff M.J., Wang P., and Casinelli J.P., 1993, \apj~419, 813
\reference{maz1998} Mazzotta P., Mazzitelli G., Colafrancesco S., and Vittorio N., 1998, \aap Suppl. ~133, 403
\reference{mcc1994} McCaughrean M.J., and Stauffer J.R., 1994, \aj~108, 1382
\reference{mew1978} Mewe R., and Schrijver J., 1978, \aap~ 65, 99 
\reference{mil2002} Miller N.A., Casinelli J.P., Waldron W.L., MacFarlane J.J., and Cohen D.H., 2002, \apj~577, 951
\reference{nor1981} Nordsieck K.H., Cassinelli J.P, and Anderson C.M., 1981, \aap~ 248, 678
\reference{ode1993} O'dell C.R., Wen Z., and Hu X., 1993, \apj ~410, 696
\reference{owo1988} Owocki S.P., Castor J.I., and Rybicki G.B., 1988, \apj~ 335, 914
\reference{owo2001} Owocki S.P., $\&$ Cohen D.H., 2001, \apj ~559, 1108 
\reference{pal1981} Pallavicini R., Golub L., Rosner R., Vaiana G.S., Ayres T., and Linsky J.L, 1981, \apj ~248, 279
\reference{pau1994} Pauldrach A.W., Kudritzki R.P., Puls J., Butler K., and Hunsinger J., 1994. \aap~ 283, 525
\reference{pet1998} Petr M.G., Coud\'e du Foresto V., Beckwidth S.V.W., Richichi A., and McCaughrean M.J., 1986, \apj~500, 825
\reference{pre1995} Predehl P. \& Schmitt J.H.M.M., 1995, \aap ~293, 889
\reference{pul1993} Puls J., Owocki S.P., and Fullerton A.W., 1993, \aap ~279, 457
\reference{rho2001} Rho J., Corcoran M.F., Chu, Y.-H., and Reach, W.T., 2001, \apj ~562, 446
\reference{rei2000} Reiners A., Stahl O., Wolf B., Kaufer A., and Rivinius T., 2000, \aap ~363, 585  
\reference{run2002} Runacres M.C. \& Owocki S.P., 2002, \aap ~381, 1015
\reference{sav1972} Savage B.D. \& Jenkins E.P., 1972, \apj ~172, 491
\reference{sch2000} Schulz N.S., Canizares C.R., Huenemoerder D., and Lee J., 2000, \apj ~545, L135 
\reference{sch2001} Schulz N.S., Canizares C.R., Huenemoerder D., Kastner J.H., Taylor S.C., and Bergstrom E.J., 2001, \apj ~549, 441
\reference{sch2001a} Schulz N.S., Huenemoerder D., Kastner J.H., and  Lee J., 2001a, AAS ~198, 2205
\reference{sch2002} Schulz N.S., 2002, in "Winds, Bubbles, and Explosions, eds. J. Arthur and R. Dyson, Patzcuaro
\reference{sch2003} Schulz N.S., Canizares C.R., Huenemoerder D., Kastner J.H., Taylor S.C., and Bergstrom E.J., 2003, \apj, erratum submitted
\reference{sew1979} Seward F.D, Froman W.R., Giaconni R., Griffith R.E., Harnden F.R. jr., Jones C., Pye J.P., 1979, \apj~ 234, 55
\reference{smi2001} Smith R.K, Brickhouse N.S., Liedahl D.A., and Raymond J.C., 2001, \apj, ~556, 91
\reference{sno1977} Snow T.P. \& Jenkins E.B., 1977, \apj S ~33, 269
\reference{spi1978} Spitzer, L. 1978, Physical Processes in the Interstellar Medium (New York: Wiley) 
\reference{sta1993} Stahl O., Wolf B., G\"ang T., Gummersbach C., Kaufer A., Kov\'acs J., Mandel H., Szeifert T., 1993, \aap~274, L29
\reference{sta1996} Stahl O., Kaufer A., Rivinius T., Seifert T., Wolf B., G\"ang T., Gummersbach C.A., Jankovics I., Kovacs J., Mandel H., Pakull M.W., and Peitz J., 1996, \aap ~312, 539
\reference{tsu2000} Tsuboi Y., Imanishi K., Koyama K., Grosso N., and Montmerle T., 2000, \apj~532, 1089
\reference{udd2002} Ud-Doula A. \& Owocki S.P., 2002, \apj ~576, 413
\reference{wal1994} Walborn N. \& Nichols J.S., 1994, \apj ~425, L29
\reference{wal2001} Waldron, W.L., \& Cassinelli, J.P., 2001, \apj, 548, L45
\reference{wal1994} Waljeski K., Moses D., Dere K., Saba J.L.R., Strong K.T., Webb D.F., and Zarro D.M., 1994,\apj~429, 909 
\reference{wei1999} Weigelt G., Balega Y., Preibisch T., Schertl D., Sch\"oller M., and Zinnecker H., 1999, \aap~347, L15 
\reference{woj2002} Wojdowski P.S., Schulz N.S., Ishibashi K., \& Huenemoerder D., in \it High Resolution
X-ray Spectroscopy with XMM-Newton and Chandra \it, MSSL, Oct. 2002
\reference{yam1993} Yamauchi S., \& Koyama K., 1993, \apj ~404, 620
\reference{yam1996} Yamauchi S., Koyama K., Sakano M., and Okada K., 1996, \pasj~48, 719
\reference{yam1999} Yamauchi S., and Kamimura R., 1999, "Star Formation 1999", sf99.proc, 308Y
\end{references}
\end{document}